\documentclass[twocolumn,tighten]{aastex61}

\newcommand{\hr}{$^\mathrm{h}$}
\newcommand{\mn}{$^\mathrm{m}$}
\newcommand{\se}{$^\mathrm{s}$}
\newcommand{\de}{$^\circ$}
\newcommand{\am}{$^\prime$}
\newcommand{\as}{$^{\prime\prime}$}

\received{June 11, 2017}
\accepted{October 3, 2017}
\submitjournal{the Astronomical Journal}

\newcommand{\dif}{\mathrm{d}}

\newcounter{column_number}
\usepackage{url}                                                                                  
\usepackage{amsmath}                                                                                  

\shorttitle{NGC~6231: Cluster Structure}
\shortauthors{Kuhn et al.}

\begin{document}

\title{The Structure of the Young Star Cluster NGC 6231. II. Structure, Formation, and Fate}

\correspondingauthor{Michael A. Kuhn}
\email{mkuhn1@gmail.com}

\author[0000-0002-0631-7514]{Michael A. Kuhn}
\affil{Millennium Institute of Astrophysics, Vicu\~na Mackenna 4860, 7820436 Macul, Santiago, Chile}
\affiliation{Instituto de Fisica y Astronom\'{i}a, Universidad de Valpara\'{i}so, Gran Breta\~{n}a 1111, Playa Ancha, Valpara\'{i}so, Chile}

\author{Konstantin V. Getman}
\affiliation{Department of Astronomy \& Astrophysics, 525 Davey Laboratory, Pennsylvania State University, University Park, PA 16802, USA}

\author{Eric D. Feigelson}
\affiliation{Department of Astronomy \& Astrophysics, 525 Davey Laboratory, Pennsylvania State University, University Park, PA 16802, USA}
\affiliation{Millennium Institute of Astrophysics, Vicu\~na Mackenna 4860, 7820436 Macul, Santiago, Chile}

\author{Alison Sills}
\affiliation{Department of Physics \& Astronomy, McMaster University, 1280 Main Street West, Hamilton, ON L8S 4M1, Canada}

\author{Mariusz Gromadzki}
\affiliation{Warsaw University Astronomical Observatory, Al. Ujazdowskie 4, 00-478 Warszawa, Poland}
\affil{Millennium Institute of Astrophysics, Vicu\~na Mackenna 4860, 7820436 Macul, Santiago, Chile}
\affiliation{Instituto de Fisica y Astronom\'{i}a, Universidad de Valpara\'{i}so, Gran Breta\~{n}a 1111, Playa Ancha, Valpara\'{i}so, Chile}

\author{Nicol\'as Medina}
\affil{Millennium Institute of Astrophysics, Vicu\~na Mackenna 4860, 7820436 Macul, Santiago, Chile}
\affiliation{Instituto de Fisica y Astronom\'{i}a, Universidad de Valpara\'{i}so, Gran Breta\~{n}a 1111, Playa Ancha, Valpara\'{i}so, Chile}

\author{Jordanka Borissova}
\affiliation{Millennium Institute of Astrophysics, Vicu\~na Mackenna 4860, 7820436 Macul, Santiago, Chile}
\affiliation{Instituto de Fisica y Astronom\'{i}a, Universidad de Valpara\'{i}so, Gran Breta\~{n}a 1111, Playa Ancha, Valpara\'{i}so, Chile}

\author{Radostin Kurtev}
\affiliation{Millennium Institute of Astrophysics, Vicu\~na Mackenna 4860, 7820436 Macul, Santiago, Chile}
\affiliation{Instituto de Fisica y Astronom\'{i}a, Universidad de Valpara\'{i}so, Gran Breta\~{n}a 1111, Playa Ancha, Valpara\'{i}so, Chile}




\begin{abstract}
The young cluster NGC~6231 (stellar ages $\sim$2--7~Myr) is observed shortly after star-formation activity has ceased. Using the catalog of 2148 probable cluster members obtained from {\it Chandra}, VVV, and optical surveys (Paper~I), we examine the cluster's spatial structure and dynamical state. The spatial distribution of stars is remarkably well fit by an isothermal sphere with moderate elongation, while other commonly used models like Plummer spheres, multivariate normal distributions, or power-law models are poor fits. The cluster has a core radius of $1.2\pm0.1$~pc and a central density of $\sim$200~stars~pc$^{-3}$. The distribution of stars is mildly mass segregated. However, there is no radial stratification of the stars by age. Although most of the stars belong to a single cluster, a small subcluster of stars is found superimposed on the main cluster, and there are clumpy non-isotropic distributions of stars outside $\sim$4 core radii. When the size, mass, and age of NGC~6231 are compared to other young star clusters and subclusters in nearby active star-forming regions, it lies at the high-mass end of the distribution but along the same trend line. This could result from similar formation processes, possibly hierarchical cluster assembly. We argue that NGC~6231 has expanded from its initial size but that it remains gravitationally bound.
\end{abstract}

\keywords{
stars: kinematics and dynamics;
stars: massive;
stars: pre-main sequence;
stars: formation;
open clusters and associations: individual (NGC~6231);
X-rays: stars}



\section{Introduction}\label{intro.sec}

\begin{figure}[h]
\centering
\includegraphics[width=0.45\textwidth]{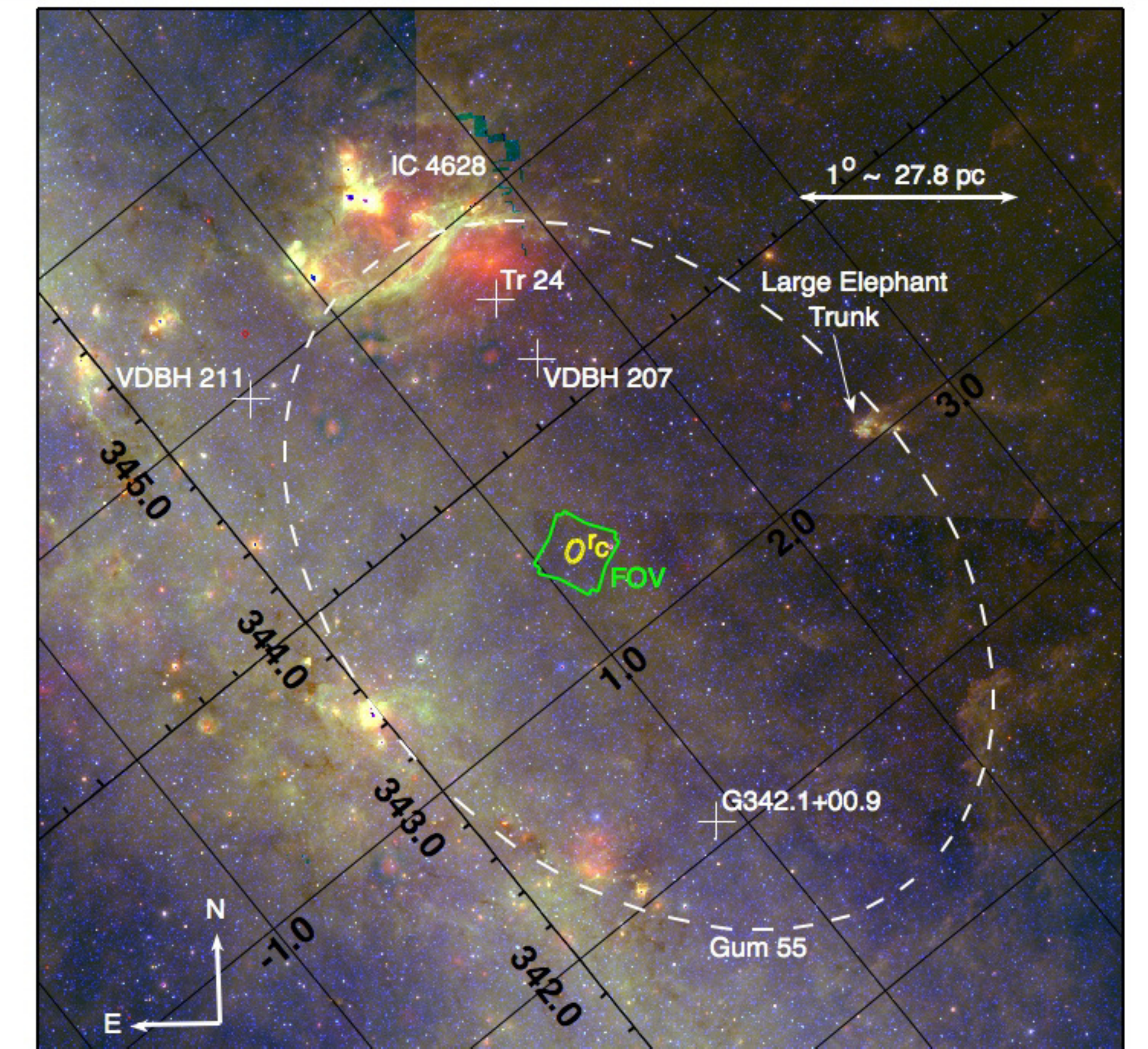} 
\caption{WISE mosaic of NGC~6231 and its surroundings in the 3.4~$\mu$m (blue), 12~$\mu$m (green), and 22~$\mu$m (red) bands, with a logarithmic color scale. The X-ray field of view is shown by the green polygon, and a yellow ellipse marks the cluster core region measured in this paper (labeled $r_c$). The Gum~55 H\,{\sc ii} region is outlined by the dashed white line, and several other nebulae and star clusters associated with Sco~OB1 are marked. The coordinate grid shows Galactic coordinates. 
\label{wise.fig}}
\end{figure}

NGC~6231 is a young star cluster (2--7~Myr) at the center of the Sco~OB1 association, on the near side of the Sagittarius spiral arm ($d\approx1.59$~kpc). This complex is $\sim\!\!1.\!\!^\circ2$ above the Galactic Plane, projected in front of the Southern Bar of the Milky Way Galaxy, so the line of sight to the cluster is very complex, with numerous field stars. The basic geometry of NGC~6231 and its environs is shown in the mid-infrared image from the WISE all-sky data release \citep{2012yCat.2311....0C} presented in Figure~\ref{wise.fig}. NGC~6231 has a substantial population of O-type stars, which power the H\,{\sc ii} region Gum~55, covering several square degrees on the sky as shown on the figure. The cluster is larger than the field of view of the {\it Chandra X-ray Observatory} (CXO) (outlined in green) that we discuss in this paper. However, like many other very young clusters \citep[cf.\ the MYStIX study;][]{2013ApJS..209...26F,2017arXiv1704.08115F}, a significant fraction of cluster members are concentrated in a dense central region that is the focus of our investigation -- the cluster core is shown by the yellow ellipse. 

The molecular cloud from which NGC~6231 formed has dissipated, but molecular clouds still exist around the periphery of the H\,{\sc ii} region, including the Large Elephant Trunk to the north-west and IC~4628 to the north-east (visible in the WISE image). The lack of molecular cloud material implies that star-formation has ceased in NGC~6231 itself, but these nearby regions may be sites of ongoing star-formation, possibly triggered by NGC~6231 \citep{2008hsf2.book..401R}. In addition to the cluster members of NGC~6231, O, B, and pre-main-sequence stars are distributed throughout the Sco OB1 association, many of which are part of a loose subcluster, Tr~24, centered $1.\!\!^\circ2$ north-east of NGC~6231 \citep{1991A&AS...90..195P}.  

The large young stellar population in NGC~6231 makes it an ideal testbed for early cluster evolution at a time just after star formation has finished. The star cluster has lost its gas, placing it at a critical stage in its evolution, where it may either disperse as an unbound association or remain as a bound open cluster. The molecular clouds that give rise to clusters like NGC~6231 are depleted both through conversion of gas to stars and by dispersal of the cloud, with dispersal accounting for most of the cloud material \citep{2003ARA&A..41...57L}. There are a variety of reasons clouds are dispersed, including ultraviolet radiation pressure \citep{2008MNRAS.391....2D,2013MNRAS.430..234D} and stellar winds from O stars \citep{2011ApJS..194...15T}, outflows from low-mass stars \citep{2006ApJ...640L.187L}, and supernovae \citep{2013MNRAS.432..455D}. 

NGC~6231 has likely had at least one supernova explosion occur in the cluster 3~Myr ago, producing the run-away high-mass X-ray binary (HMXB) HD~153919 $\sim$4$^\circ$ from the cluster \citep{2001A&A...370..170A}. It is probable that the progenitor of this supernova was one of the first O stars formed in NGC 6231 because the cluster has $\sim$20 main-sequence members with masses above the $\sim$8~$M_\odot$ limit for supernova explosions and only one Wolf-Rayet star in the field of view \citep[][and references therein]{KMG17}.

Stars born in massive star-forming complexes are often initially gravitationally bound to the complex \citep[e.g.,][]{2014A&A...563A..94J,2015ApJ...812..131K,2015A&A...578A..35M,2016A&A...588A.123R}. However, gravitationally bound groups of stars can be disrupted by cloud dispersal \citep{1978A&A....70...57T,1980ApJ...235..986H,1984ApJ...285..141L} or by tidal interactions with external giant molecular clouds \citep{2012MNRAS.426.3008K}. If groups of stars are formed with sufficiently high star-formation efficiencies and are sufficiently massive, they may survive as open clusters. Nevertheless, most groups will disperse as unbound associations \citep{2003ARA&A..41...57L}, and even surviving open clusters may lose a large fraction of their stars \citep{2001MNRAS.321..699K}. Investigation into cluster survival has focused on both star-formation efficiency and cluster structure \citep[e.g.,][]{2001MNRAS.321..699K,2006MNRAS.369L...9B,2006MNRAS.373..752G,2009A&A...498L..37P,2011A&A...536A..90P,2013A&A...555A.135P,2013A&A...559A..38P,2014ApJ...794..147P,2015A&A...576A..28P,2012MNRAS.426.3008K,2015MNRAS.448.2504G,2017A&A...597A..28B}.

It has been previously hypothesized that NGC~6231 is in the process of evolving into an unbound association \citep[e.g.,][]{2015MNRAS.448.1687S}. We will use the structural properties of NGC~6231, revealed by a more-complete census of its cluster members, to help to address the fate of this young star cluster.

\defcitealias{KMG17}{Paper~I}
This paper is the second in a two-paper investigation of NGC~6231 using observations from the {\it Chandra X-ray Observatory} (CXO) and the VISTA Variables in the V\'ia Lact\'ea (VVV) survey \citep{2010NewA...15..433M}. The first paper \citep[viz.,][henceforth Paper~I]{KMG17} obtains a new census of the stellar population while this paper addresses cluster structure. 
Section~\ref{cat.sec} describes the membership catalogs from \citetalias{KMG17}, and Section~\ref{cluster_properties.sec} provide estimates of total populations from these incomplete catalogs. Section~\ref{morph.sec} discusses the spatial distribution of cluster members, and Section~\ref{mixture.sec} models their surface density distribution. Section~\ref{seg.sec} studies mass segregation in NGC~6231. Section~\ref{age.sec} investigates gradients in stellar ages. Section~\ref{largescale.sec} examines the spatial distribution of stars outside the {\it Chandra} field of view. Section~\ref{discussion.sec} discusses the implications of the observed cluster properties on the cluster's formation and fate. And, Section~\ref{conclusions.sec} provides the conclusions.

Many of the data reduction and analysis methods used in this study were developed for the Massive Young Star-Forming Complex Study in Infrared and X-ray (MYStIX) project \citep[][and references therein]{2013ApJS..209...26F} -- a comparative study of 20 young star clusters in nearby massive star-forming regions. These include methods for identifying and modeling subclusters of stars \citep{2014ApJ...787..107K}, methods for analyzing the intrinsic densities of stars in clusters \citep{2015ApJ...802...60K}, methods for analyzing mass segregation (M.~A.~Kuhn et al., in preparation), and methods for identifying gradients in stellar ages \citep{2014ApJ...787..109G,2014ApJ...787..108G}. Many alternative methods for these types of analysis are found in the literature. However, by using MYStIX methods, it is easier to make comparisons between different young star clusters observed with {\it Chandra}.

For this study we adopt a distance of 1.59~pc for the cluster and a median age of 3.2~Myr for the stellar population, but with star-formation activity going back at least 6.4~Myr \citepalias{KMG17}. 

\section{Catalogs of Probable Cluster Members \label{cat.sec}}

The study of cluster structure is largely based on the catalog of 2148 probable cluster members in the {\it Chandra} field of view from \citetalias{KMG17}. This CXOVVV catalog includes X-ray sources, classified based on X-ray properties and optical/near-infrared counterparts, spectroscopic OB stars obtained from the literature, and near-infrared variables from the VVV survey. The initial X-ray catalog is estimated to include 130$\pm$30 unrelated fields stars. Likely field stars were filtered out on optical color-magnitude diagrams if they appeared too far above or below the locus of cluster members; however, some may remain in the final sample. 

Most of the sources in the CXOVVV catalog are low-mass pre-main-sequence stars \citepalias{KMG17}. The known high-mass stellar content includes 1 Wolf-Rayet star, 13 O-type stars, and 82 B-type stars. In \citetalias{KMG17}, stellar masses are estimated using near-infrared $JHK_s$ photometry assuming \citet{2000A&A...358..593S} pre-main-sequence evolutionary models or using early-type stars' spectral classifications. \citet{2010ApJ...725.2485K} found that a similar mass-estimation method for pre-main-sequence stars in Taurus yielded typical errors of 0.15--0.30~dex. Stellar age estimates are obtained with two methods: the $V$ vs. $V-I$ color-magnitude diagram ($Age_{VI}$) and relations between X-ray luminosity and $J$-band luminosity \citep[$Age_{JX}$;][]{2014ApJ...787..108G}. 

\subsection{Outside the {\it Chandra} Field of View}\label{outside.sec}

Methods for selection of cluster members with only optical/near-infrared data, available in the region outside the {\it Chandra} field of view, yield samples with much higher rates of non-member contamination. However, even these catalogs may be useful for visualizing spatial clustering on larger spatial scales. We use both variability and color-magnitude diagrams for selection. 

A total of 295 near-infrared variables are found in a $2.\!\!^\circ3\times1.\!\!^\circ5$ box surrounding NGC~6231 (VVV tiles ``d148'' and ``d110''). Near-infrared variability in pre-main-sequence stars may be produced by star spots, accretion from a circumstellar disk, and variable extinction from the circumstellar disk \citep[e.g.,][]{1945ApJ...102..168J,1994AJ....108.1906H}.  In a representative study of high-amplitude variables in the VVV survey by \citet{2016arXiv160206267C,2016arXiv160206269C}, it was estimated that approximately 50\% of near-infrared variables were pre-main-sequence stars. 

\citet{2016arXiv160708860D} note that candidate O--B and A--F stars can be selected using the optical photometry alone, and that many of these stars will be cluster members. We use photometry from the VST Photometric H$\alpha$ Survey of the Southern Galactic Plane and Bulge \citep[VPHAS+;][]{2014MNRAS.440.2036D} to identify candidate stars with spectral types of F or earlier in the portion of the Sco~OB1 survey covered by VPHAS+. The magnitude and color cuts, $g<15.5$~mag and $g-i<$1.5~mag, are designed to approximate the selection rules from \citet{2016arXiv160708860D}, which used the Johnson-Cousins $V$ and $I$ bands instead. From the spatial distribution of these stars (\S\ref{obaf}), it is clear that there is also a large unclustered population, which are likely to be field stars.

 \begin{figure*}[t!]
\centering
\includegraphics[width=1.0\textwidth]{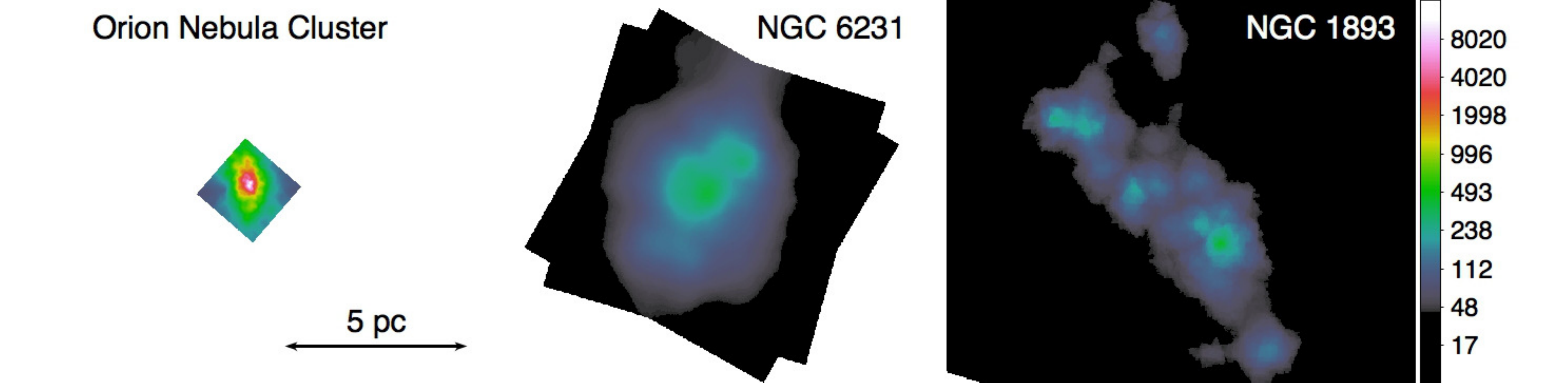} 
\caption{Maps of surface density [stars~pc$^{-2}$] of X-ray selected stars in NGC~6231 (center) compared to the Orion Nebula Cluster (left) and NGC~1893 (right). All three regions are shown to the same physical scale (a 5-pc length scale is shown) and have been corrected for differences in X-ray sensitivity (a color-bar shows densities scaled to the full IMF). The fields are oriented with north up and east to the left.
\label{dens.fig}}
\end{figure*}

\section{Intrinsic Stellar Population \label{cluster_properties.sec}}

Even with the deep {\it Chandra} exposure, most stellar-mass cluster members are not detected.
The X-ray catalog is only complete for sources with X-ray photon fluxes greater than $\log F_\mathrm{photon}=-5.95$ [photon~s$^{-1}$~cm$^{-2}$] in the 0.5--8.0~keV band. For pre-main-sequence stars, there is a positive correlation between stellar mass $M$ and X-ray luminosity $L_X$ \citep[e.g.,][]{2007A&A...468..425T}, so higher-mass pre-main-sequence stars are more likely to be detected while lower-mass pre-main-sequence stars are less likely to be detected. 

In \citetalias{KMG17}, we use both the initial mass function (IMF) and X-ray Luminosity Function (XLF) to extrapolate the number of stars missed in the observations \citep[e.g.,][]{2015ApJ...802...60K}. The IMF method from \citetalias{KMG17} gives a total of 5700$\pm$250 stars projected within the {\it Chandra} ACIS-I field of view, using the masses calculated with a 3.2-Myr isochrone. However, if an older age of 6.4~Myr were assumed, the estimated number of stars would increase to 7500$\pm$360. The XLF analysis gives 6000$\pm$530 stars assuming that the change in shape of the XLF is small during the first 5~Myr. The uncertainties reported above reflect statistical errors alone. These are calculated using bootstrap resampling with replacement (1000 draws), where normally distributed random errors were added to $\log M$ or $\log L_X$ for each draw -- a factor of 2 for stellar mass or the reported uncertainty for X-ray luminosity. 
 
Systematic uncertainties due to stellar-mass estimation and assumptions about the IMF and XLF are difficult to quantify and may be larger than the statistical uncertainties. For example, \citet[][their Figure~4]{2015ApJ...802...60K} compared IMF and XLF estimates for the MYStIX subclusters and found discrepancies of $\sim$0.25~dex. This may be regarded as a pessimistic estimate of the systematic uncertainty on number of stars in NGC~6231. However, the estimates for NGC~6231 may be more accurate than for MYStIX subclusters because of NGC~6231's higher number of observed stars, its lower extinction and lack of strong variation in extinction, and its lack of infrared nebulosity. The IMF and XLF estimates of 5700$\pm$250 and 6000$\pm$530 stars in NGC~6231 agree within their estimated statistical uncertainties.

For an analysis of cluster structure, it is essential that inhomogeneities in detector sensitivity not affect star counts.  {\it Chandra}'s sensitivity is greatest near the optical axis of the telescope, leading to a larger number of faint sources being detected near the cluster center \citep[][]{2011ApJS..194....2B}. Thus, we remove sources with X-ray fluxes below the photon-flux completeness limit from the study, as has been done for previous work \citep[e.g.,][]{2011ApJS..194....9F,2014ApJ...787..107K}. This leaves 826 X-ray sources, combined with all known O and B-type stars, for a total sample of 885 objects out of a total of 2148 probable cluster members. If we assume that the total number of stars projected in the field of view is 5700, a correction factor of 6.5 would need to be applied to any star counts or surface densities measured from this sample to obtain an intrinsic astrophysical value.  (The correction factor would be 8.5 if the older age were assumed.)

The 50\% mass completeness limit found by the IMF analysis is 0.5~$M_\odot$ \citepalias{KMG17}. Source detection probability rolls off gradually as a function of stellar mass, due to the scatter in the $L_X$--$M$ relation, so the limit we report for mass is where a pre-main-sequence stars has a 50\% chance of being detected; but most pre-main-sequence stars and OB stars above this limit will be detected. Late-B and A stars are expected to be missing from X-ray surveys; however, some of these stars do have X-ray emitting pre-main-sequence companions \citepalias{KMG17}.

\section{Cluster Morphology \label{morph.sec}}

The projected surface density of stars in NGC~6231, normalized to the full 5700-member population, is displayed in Figure~\ref{dens.fig} (center panel), along with similarly normalized stellar surface-density maps of two MYStIX star-forming regions, the Orion Nebula Cluster (left panel) and NGC~1893 (right panel), obtained by \citet{2015ApJ...802...60K}. 
These maps were generated by adaptively smoothing the spatial point patterns of stars using the algorithm {\it adaptive.smoothing} from the R software package {\it spatstat} \citep{baddeley2005spatstat,baddeley2015spatial}. This algorithm subdivides the field of view using the Voronoi tessellation, using a fraction $f$ ($=5\%$) of points to generate the tiles and a fraction $1-f$ ($=95\%$) of points to estimate surface density in each tile. This procedure is repeated 500 times, and the results are averaged to produce the smoothed maps. The values in the surface-density maps are then multiplied by correction factors (\S\ref{cluster_properties.sec}) to estimate intrinsic surface density.\footnote{Use of the correction factor requires the assumption that X-ray properties of stars are not correlated with position. This assumption does not necessarily hold because the X-ray photon flux is correlated with stellar mass, and the cluster is likely to be mass segregated. Nevertheless, the star-counts are dominated by lower mass stars which are not strongly mass segregated.} A complete description of how the maps were generated for the two MYStIX regions is given by \citet{2015ApJ...802...60K}. 

The three clusters in Figure~\ref{dens.fig} are shown using the same spatial scale (a 5~kpc arrow is shown) and same surface-density scale (indicated by the color bar) to allow their morphologies to be directly compared. Although all three regions have similar ages and numbers of stars (approximately 2.5~Myr and 2600 stars for the Orion Nebula, 3.2~Myr and 5700 stars for NGC~6231, and 2.6~Myr and 4600 stars for NGC~1893), there are significant differences in cluster structure. The Orion Nebula Cluster is surrounded by a dense molecular cloud, but the center of the cluster is partially evacuated; in contrast, the bubbles around NGC~6231 and NGC~1893 are much larger. NGC~6231 is much less dense than the Orion Nebula Cluster, with a central surface density of $\sim$300~stars~pc$^{-2}$ compared to $>$10,000~stars~pc$^{-2}$, and has a larger physical size. NGC~6231 is similar in size to NGC~1893; however, NGC~1893 is significantly more clumpy and elongated, being divided into $\sim$10 subclusters. The individual subclusters in NGC~1893 also appear physically smaller than the NGC~6231 cluster, although they have similar peak surface densities. 

When compared to 17 star-forming regions from MYStIX, shown by \citet[][their Figure~5]{2015ApJ...802...60K}, NGC~6231 appears atypical in having a simple structure with a much larger size than the typical subcluster, rather than a collection of denser subclusters. This suggests that NGC~6231 has expanded considerably from its initial size. NGC~6231 may be most similar to NGC~2244, an expanded cluster of stars that is part of the Rosette Nebula star-forming region. However, NGC~2244 in Rosette does not show mass segregation \citep{2008ApJ...675..464W} whereas NGC 6231 does (\S \ref{seg.sec}).

\section{Surface Density Models \label{mixture.sec}}

Several families of spherically symmetric models have been used to fit the (surface) density profiles of star clusters. Models include the isothermal sphere, the King profile, and the Plummer Sphere, which all represent approximations to density profiles of virilized, gravitationally bound groups of stars in a quasi-equilibrium state \citep{2008gady.book.....B}. None of the cluster profiles would necessarily be expected to provide a good model for young clusters ($\mathrm{age}<5$~Myr), which are not expected to be in dynamical equilibrium due to their young age and would show an imprint from formation in their natal molecular clouds. The lack of kinematic data for cluster members means that we cannot assess the thermodynamic state of the star cluster, so we use these models as empirical descriptions of the surface density.

Other possible models include multivariate normal distributions and a power-law radial-density gradients. If stars have a Maxwell-Boltzmann velocity distribution and are allowed to freely expand from a point, then their resulting spatial distribution would be a multivariate normal distribution. A radial power-law surface-density distribution has also been proposed as a possible model for young star clusters by \citet{2004MNRAS.348..589C}. The power-law model is scale invariant, unlike models that require a critical length scale $r_c$. 

We fit the projected spatial distribution of stars using several functional forms based on the above models. \begin{eqnarray}
\label{hubble.eq}
\mathrm{Hubble~model}~~&~\Sigma(R) =\Sigma_0\left[1+(R/r_c)^2\right]^{-1} \\
\label{plummer.eq}
\mathrm{Plummer~sphere}~ &~\Sigma(R) =\Sigma_0\left[1+(R/r_c)^2\right]^{-5/2} \\
\label{normal.eq}
\mathrm{Normal~distribution}~ &~\Sigma(R) =\Sigma_0 \exp\left(-R^2\middle/2r_c^2\right)\\
\label{powermod.eq}
\mathrm{Power~law}~ &~\Sigma(R) =AR^{\alpha},~\mathrm{with}~\alpha>-2
\end{eqnarray}
In these equations $R$ is the projected distance from the center of the cluster, $\Sigma_0$ is the surface density at the center of the cluster, $r_c$ is a characteristic radius called the ``core radius'' in the case of the Hubble model and the Plummer sphere, $A$ is a scaling constant, and $\alpha$ is a power-law index. 
The isothermal sphere has both singular (density diverges at $R=0$) and non-singular solutions.
The non-singular solution can be approximated out to several core radii by the Hubble model \citep{2008gady.book.....B} above.  
However, for $R\gg r_c$, $\Sigma(R)$ asymptotically approaches a power-law dependence on $R$, with an index of $\alpha=-1$, not $\alpha=-2$ as suggested by the Hubble model. The singular isothermal sphere is a power-law model with $\alpha=-1$.

Often the distributions of stars show elongation; this includes the Orion Nebula Cluster \citep{1998ApJ...492..540H} and many of the subclusters in the MYStIX star-forming regions \citep{2014ApJ...787..107K}. To account for elongation of subclusters, \citet{2014ApJ...787..107K} introduced additional parameters to the model for ellipticity, $\epsilon$, and orientation, $\phi$, in projection on the sky. Equations~\ref{hubble.eq}--\ref{powermod.eq} can be redefined to describe ellipsoidal distributions if we redefine $R$ as 
\begin{multline}
R =  
\left| \left[ \begin{array}{cc}
(1-\epsilon)^{-1/2}\cos \phi & (\epsilon-1)^{1/2}\sin \phi \\
(1-\epsilon)^{-1/2}\sin \phi& (1-\epsilon)^{1/2}\cos \phi \\
\end{array} \right]
\left[ 
\begin{array}{c}
\Delta x\\ \Delta y \\
\end{array}
\right]
\right|,
\end{multline}
where $\Delta x$ and $\Delta y$ are the distances of a point from the cluster center along $x$- and $y$-axes. Following \citet{2014ApJ...787..107K}, we refer to the distribution described by the transformed version of Equation~\ref{hubble.eq} as the ``isothermal ellipsoid,'' although this is an empirical model rather than one derived from physics. 

We use the maximum likelihood method to fit these models to the spatial distributions of cluster members, using software provided by \citet[][their Appendix~A]{2014ApJ...787..107K} written in the R programming language. The log-likelihood is 
\begin{equation}\label{likelihood}
\mathcal{L}(\theta;\mathbf{X})=\sum_{i=1}^{N} \ln \Sigma_{\theta}(\mathbf{r}_i)-\int_W \Sigma_{\theta}(\mathbf{r}^\prime)\dif \mathbf{r}^\prime,
\end{equation}
where $\mathbf{X}=\{\mathbf{r}_i\}$ is the set of points, $\theta$ are the model parameters, and $W$ is the field of view. The Nelder-Mead algorithm is used to optimize $\mathcal{L}$, while the Hessian matrix of $\mathcal{L}$ at the maximum likelihood parameters $\hat{\theta}$ is used to estimate uncertainty in model parameters.

The resulting cluster parameters include coordinates of the cluster center, the core radius $r_c$, the ellipticity $\epsilon$ and orientation $\phi$ of the cluster, and the central density $\Sigma_0$ for the isothermal ellipsoid, the Plummer ellipsoid, and the multivariate normal distribution. For the singular isothermal sphere and power-law model, the parameter $r_c$ is not needed, and for the power-law model the index $\alpha$ is also fit. 

\subsection{Single-Cluster Models}\label{single.sec}

\begin{figure*}[h]
\centering
\includegraphics[width=0.925\textwidth]{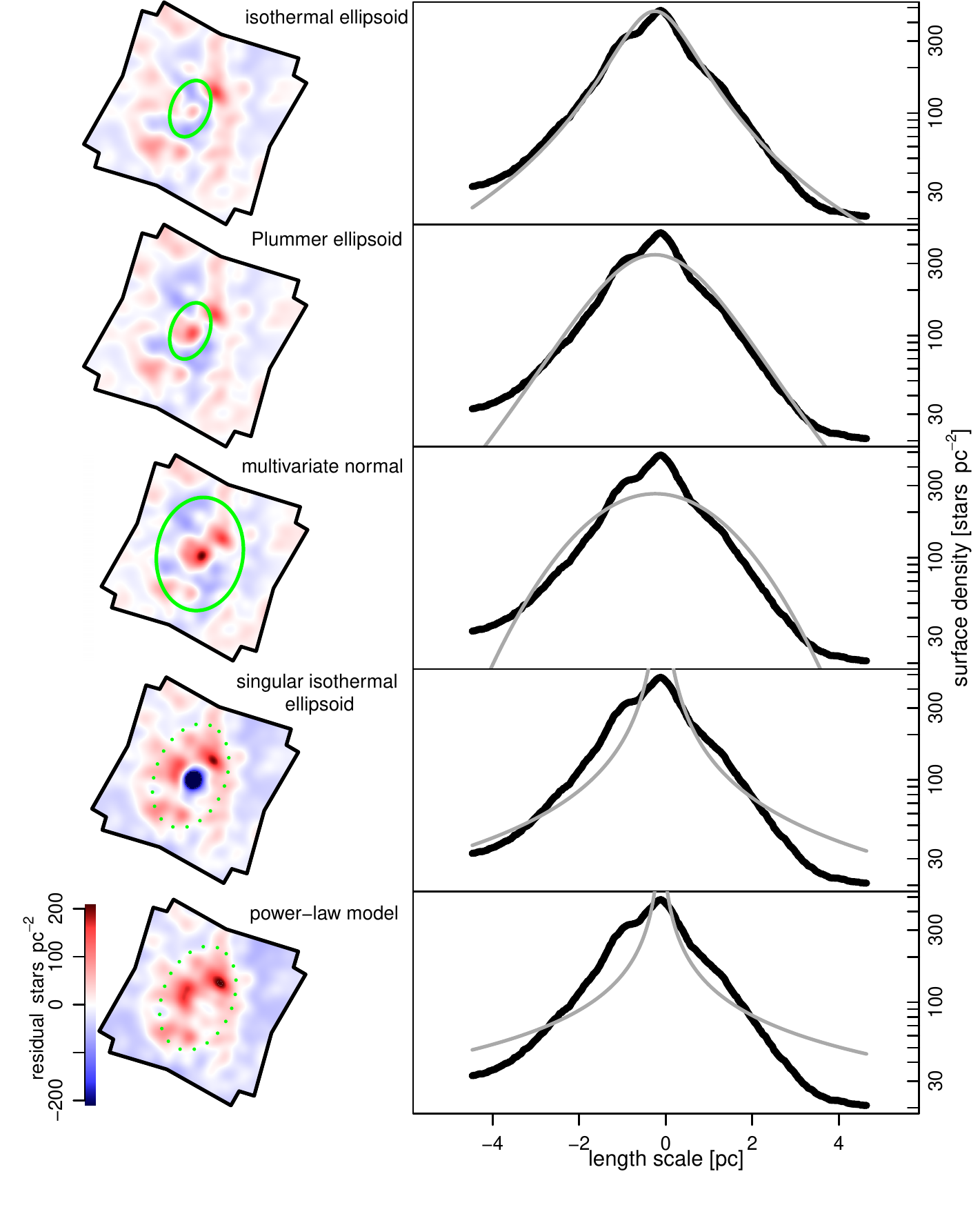} 
\caption{
Residual maps (left column) and surface density profiles (right column) for the 5 cluster models.  Left: Residuals were calculated using a Gaussian kernel with $\sigma_\mathrm{BW}=0.3$~pc. Negative residuals are blue and positive residuals are red. Possible subclusters appear as deep red spots. The green ellipses indicate model ellipticity and orientation, with solid lines indicating core radius $r_c$ and dashed lines showing a representative radius. Right: Surface density along a constant constant declination ($\delta=-41$\de 50\am00\as) for the adaptively smoothed data (black line) and the model (gray line).
\label{cuts.fig}}
\end{figure*}

The distribution of stars in NGC~6231 was fit with the isothermal ellipsoid, the Plummer ellipsoid, the multivariate normal, the singular isothermal ellipsoid, and the power-law models (Figure~\ref{cuts.fig}). We omit the King model, a modification of the isothermal sphere with an outer truncation radius, because the cluster is truncated by the field of view. In the figure, the panels in the left column show green ellipses, representing model parameters, overplotted on spatial maps of fit residuals. These ellipses indicate where the cluster is centered, the ellipticity and orientation of the model, and, for models with a characteristic radius, the size of the cluster core. The panels in the right column show a one-dimensional (constant declination) slice through the best-fit parametric models (gray lines) and the adaptively smoothed data (black lines). The $y$-axes are plotted with logarithmic values to allow a high contrast in surface density to be shown. 

Goodnesses of fit can be evaluated using the residual maps (Figure~\ref{cuts.fig}, left column). The residual maps are produced by the function {\it diagnose.ppm} in {\it spatstat}. This tool smooths both the model and the points with a kernel ($\sigma=0.3$~pc), and the residual ($\mathrm{model}-\mathrm{points}$) map is shown by the color scale in units of stars~pc$^{-2}$. The mathematical foundations for this analysis are given by \citet{Baddeley05,Baddeley08}. In these residual maps, a good fit is indicated by a residual value of 0 (white), while positive and negative values (red or blue) indicated differences between the data and the model. Thus, patches of dark red may represent subclusters of stars not accounted for by the models. For the singular isothermal ellipsoid and the power-law models, which diverge at $R=0$, the number of stars remains finite when $\alpha>-2$, so the kernel will smooth over the divergence at the center.

A description of the best fits is given below.
\begin{description}
\item[Isothermal ellipsoid] There is close agreement between the model and the adaptively smoothed data in both the cluster center and in the outer region of the cluster, out to $R\sim4$~pc (most of the field of view). In particular, in the cluster wings, the slope of the model is a good match to the smoothed data. Residuals are small ($\pm$25~residual~stars~pc$^{-2}$ which is $<$10\% of the peak surface density) except for a peak of 120~residual~stars~pc$^{-2}$ to the north-west of the cluster center. The deviation in the outer part of the cluster may be partially due to the inadequacies of the Hubble model as an approximation for the isothermal sphere, but the numbers of stars in these regions are small, so the differences could also be due to stochastic fluctuations in counting statistics.
\item[Plummer ellipsoid] There is a significant difference between the model and the adaptively smoothed data at the cluster center. This arises because the Plummer model has a flatter core region than the isothermal ellipsoid, as well as having a steeper decrease in surface density with radius outside the cluster core. The underestimate in number of stars in the cluster center can be seen as a positive residual at this location in the smoothed map.
\item[Multivariate normal] The central density is significantly underestimated, and the rapidly decreasing wings of the normal distribution do not match the shape of the smoothed data, which is more gradually decreasing. The 1-sigma ellipse of the distribution is significantly larger than the core radius found by the isothermal ellipsoid and the Plummer ellipsoid models because the multivariate normal distribution must be enlarged to fit the broad wings of the distribution of stars. 
\item[Singular isothermal ellipsoid] The cusp at the center of this model strongly overestimates the number of stars at the cluster center. Furthermore, the slope (given by $\alpha=-1$) is too flat in the outer regions of the model producing a strong negative residual surrounded by a ring of positive residuals.  
\item[Power-law model] Allowing $\alpha$ to be a free parameter decreases the number of stars in the central cusp, decreasing the negative residual. Nevertheless, the power-law index of the model ($\alpha=0.7$) is even flatter in the outer regions of the cluster than the singular isothermal ellipsoid model, poorly matching the adaptively smoothed data. 
\end{description}

Thus, the isothermal ellipsoid model provides a remarkably good empirical approximation of the observed surface density distribution, while all other models are clearly inadequate. This is good motivation for the use of the isothermal ellipsoid form for analysis of NGC~6231. The good fit found using the isothermal ellipsoid model is consistent with the results from the MYStIX clusters RCW~38, the Flame Nebula Cluster, and M~17, even though these clusters are much denser \citep{2014ApJ...787..107K}.  The physical cluster parameters based on the isothermal ellipsoid model are provided in Table~\ref{cluster2.tab}, labeled ``Model~1.''

Whether or not young star clusters have a critical length scale has been an open question \citep[e.g.,][]{2001AJ....121.1507E,2004MNRAS.348..589C}. For NGC~6231, the core radius $r_c$ for the isothermal ellipsoid model appears to be such a length scale. The core radius, $r_c=1.2\pm0.1$, is inconsistent with a value of 0. In addition, both scale invariant models---the singular isothermal ellipsoid or the power-law model---poorly fit the data.

When the distribution of stars is fit with only one cluster component, the residual maps show an overdensity of stars north west of the cluster center, at coordinates 16\hr54\mn00\se.6 $-$41\de48\am07\as. This residual represents anisotropic structure that could indicate another subcluster. We investigate this possibility quantitatively in Section~\ref{two.sec}.

For the model fitting presented above, all stars of different masses are all treated equally. To investigate whether the combination of stars of different masses affects the functional form of the models, we redo model fitting for 768 low-mass stars, excluding stars with $M\ge3$~$M_\odot$.  The results are nearly the same as when using the full mass range. The isothermal ellipsoid provides the best fit to both the cluster core and cluster wings, while the normal distribution underestimates the number of stars in the core and the scale-invariant models both yield results that are too cuspy in the center and have slopes that are too flat in the outer regions. However, for the smaller sample, it is more difficult to distinguish between the Plummer ellipsoid and the isothermal ellipsoid.  A residual corresponding to the candidate subcluster is still apparent for all models.

\subsection{Two-Subcluster Model}\label{two.sec}

Beyond the single-cluster model, it is possible to model multiple cluster components using a statistical method known as mixture models (see review by Kuhn \& Feigelson, 2017). A mixture model is a probabilistic model in which the probability density function (e.g., surface density) for a set of points is composed of the sum of multiple probability density functions for subpopulations (e.g. subclusters of stars). 

For NGC~6231, each component has the form of an isothermal ellipsoid model, which may be used to model different groups of stars in the region. This method of cluster analysis was used by \citet{2014ApJ...787..107K} to identify subclusters in MYStIX. For NGC~6231, two subclusters are suspected: the main cluster and the possible subcluster to the north-west -- the smaller subcluster is designated Subcl.~A following the notation of \citet{2014ApJ...787..107K}.

Determining whether a population of stars is one or more clusters can be viewed as a model selection problem. Penalized likelihood methods are commonly used for mixture model problems, including the Akaike informaion criterion \citep[AIC;][]{akaike1974new} and Bayesian information criterion \citep[BIC;][]{schwarz1978estimating}, defined by the formulas
\begin{eqnarray}
\mathrm{AIC}&=&-2\mathcal{L}+2k,\\
\mathrm{BIC}&=&-2\mathcal{L}+k\ln n,
\end{eqnarray}
where $\mathcal{L}$ is the log-likelihood, $k$ is the number of parameters in the model, and $n$ is the number of points. For the isothermal ellipsoid clusters, $k$ is equal to 6 times the number of clusters, and we select the $k$ that minimizes the AIC or BIC. The BIC generally favors simpler models than the AIC due to its larger penalty for the inclusion of additional parameters.  The choice of AIC, BIC, or other model selection approaches has been widely debated \citep{Lahiri01,Burnham02,Konishi08}.

The best-fit two-component model includes a main cluster, with properties similar to the single-cluster model, and a small subcluster, coincident with the over-density of stars to the north-east of the main cluster. The log-likelihood increases from 1599 to 1618 with the addition of the subcluster (log likelihood will always be greater for the model with more components). Both the BIC and AIC favor the model with two components. The difference in BIC values, $\Delta BIC = 36$, is far above the threshold $\Delta BIC \simeq 8-10$ for confident preference of the 2-cluster model \citep{KassRaftery95}. Integrated over the entire field of view, the ratio of the number of stars in the main cluster to the number of stars in the subcluster is 24:1. 

\begin{figure*}[t]
\centering
\includegraphics[width=1.0\textwidth]{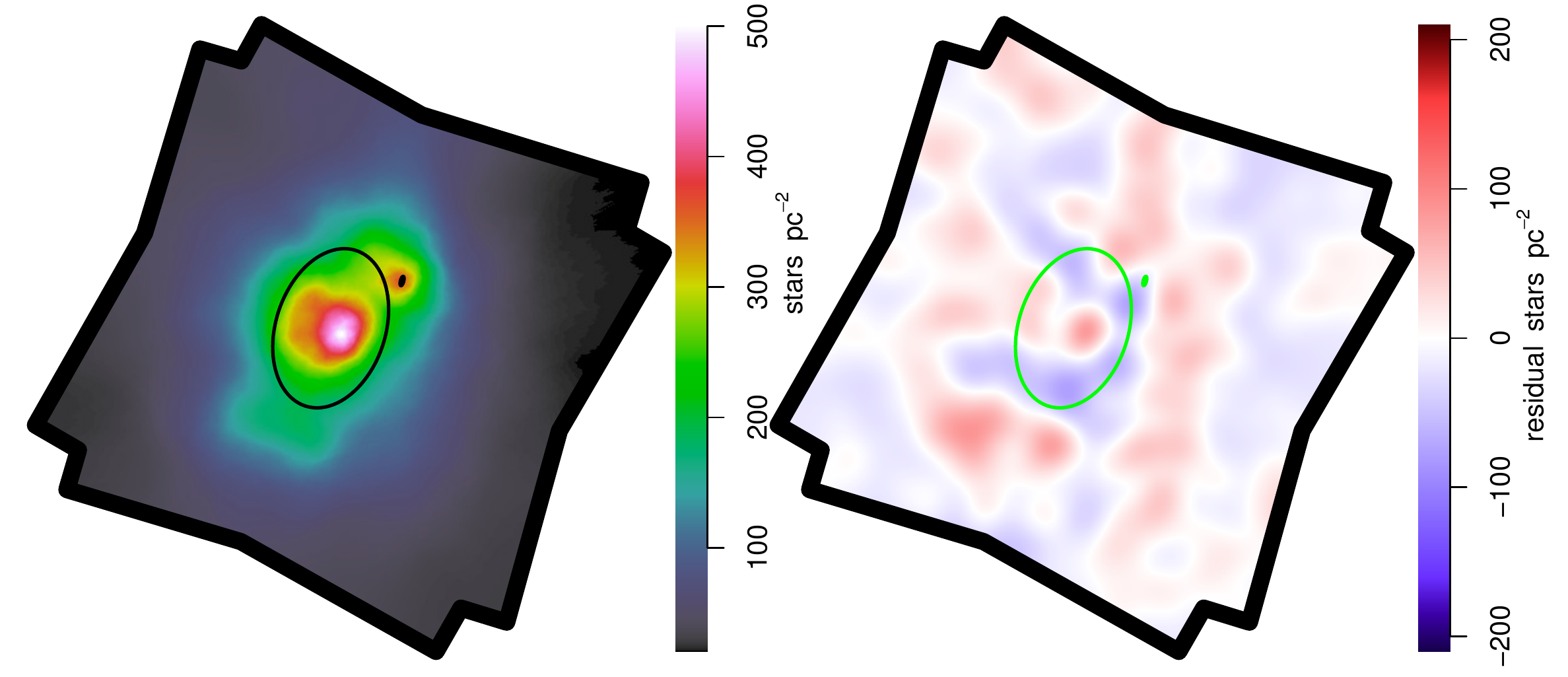} 
\caption{ Left: Surface density estimate for stars in NGC 6231 with the cluster-core contours overplotted (black ellipses). The two ellipses correspond to the two-component model in Table~\ref{cluster2.tab}. Right: The residual plot for the two-component model. The color-scale is the same as in Figure~\ref{cuts.fig} (bottom row), but the positive residual to the north-west of the main cluster is gone. 
 \label{two_comp_model.fig}}
\end{figure*}

The cluster and subcluster are shown in Figure~\ref{two_comp_model.fig}, where the elliptical contours marking the cores for each component are plotted on an adaptively smoothed surface density map and on a map of smoothed residuals. The addition of the second cluster significantly decreases the amplitude of the residuals. The core radius of the subcluster is clearly much smaller than that of the main cluster (although not too dissimilar from many subclusters found in MYStIX star-forming regions). 

\subsection{Properties of the Main Cluster and Subcluster}

Table~\ref{cluster2.tab} presents the best-fit models for the single-component model and the two-components model. This includes parameters and uncertainties estimated directly from model fitting, including component centers, core radius, ellipticity, and orientation. Numbers of stars and central surface and volume density also include corrections for incompleteness. For the Hubble model, the central volume density, $\rho_0$ is related to the central surface density, $\Sigma_0$, by the relation $\rho_0 = \Sigma_0 / 2 r_c$. This equation will also apply to the ellipsoidal model if we assume that the three-dimensional ellipsoid has a harmonic-mean core radius of $r_c$.

The mixture model provides precise and reliable celestial coordinates of the centers of the main cluster and subcluster (Table~\ref{cluster2.tab}). The modeling finds the centroid of the star counts belonging to each cluster, correcting for truncation of the field of view and overlapping of the clusters. For the rest of the work we will use the coordinates 16\hr54\mn15.\se9~$-$41\de49\am59\as\ as the definition of the center of the main cluster. This is offset by $\sim$0.5~arcmin to the east of the location of the peak of the adaptively-smoothed surface-density map.\footnote{ The difference between the centers is an indication of the slight asymmetry in the distribution of stars. The precise position of the peak density may depend on the smoothing algorithm that is used.}  

The subcluster is located at 16\hr54\mn1.6\se~$-$41\de48\am13\as. 
Although there is considerable uncertainty in some subcluster properties from the model fit, the number of stars in the subcluster, integrated over the whole field of view, $240\pm40$~stars, is less strongly dependent on the core-radius model parameter. The O9.7Ia+O8V system HD~152234 is projected at the location of the subcluster. However, this alignment could be coincidental due to the large number of massive stars in the cluster. 
 
Another anisotropy in the spatial distribution of stars of the main cluster is a shoulder in surface density, 0.5~pc east of the cluster peak. This can be seen in both the smoothed surface-density maps of both Figure~\ref{cuts.fig} (bottom row) and in the surface-density profile in Figure~\ref{cuts.fig} (top row). This asymmetry does not correspond to its own mode in surface density, and it could be merely a statistical fluctuation in the distribution of stars.

\subsection{Characterizations of Cluster Size and Total Mass \label{tot_mass.sec}}

Estimates of cluster radius and mass must be carefully defined for young star clusters in star-forming complexes like Sco~OB1. In theoretical studies of cluster evolution, both the cluster mass and the half-mass radius $r_\mathrm{eff}$ of a cluster (i.e.\ the radius of a sphere that encompasses half the mass of the cluster) have been important quantities for characterizing clusters. Neither of these can be directly measured for the isothermal ellipsoid model of NGC~6231 because the cluster is truncated by the field of view. Instead, we report two values that are well defined by our model: $N_{4}$, the number of stars (corrected for incompleteness) within a region 4-times larger than the cluster core, and the corresponding radius $r_4=4 r_c$. In general there is no fixed ratio between the core radius and half-mass radius, but for the clusters in the \citet{2010ARA&A..48..431P} sample, $\log (r_\mathrm{eff} / r_c) \sim 0.7\pm0.4$~dex, so $r_4$ is likely within a factor of several of the half-mass radius. For $N_{\mathrm{core}}$ and $N_{4}$, we provide the uncertainty on the number of stars within the reported radii, including only the statistical uncertainties on number of stars. For $\Sigma_0$ and $\rho_0$, uncertainty from estimation of core radius is also included.

  \startlongtable
\onecolumngrid
\newpage  
\begin{deluxetable}{ll|c|cc}
\tablecaption{Best-fit Cluster Models \label{cluster2.tab}}
\tabletypesize{\small}
\tablehead{
\colhead{} & \colhead{} &\colhead{Model 1} &  \multicolumn{2}{c}{Model 2}\\
\colhead{} &\colhead{} & \colhead{Main}  &  \multicolumn{2}{c}{~~~Main~~~~$+$~~~~Subcl. A}
}
\startdata
R.A. & (J2000) & 16 54 14.7 [0.5]&   16 54 15.9 [0.6] & 16 54 1.6 [0.1]\\
Decl. &  (J2000) & $-$41 49 47 [11]& $-$41 49 59 [12]&  $-$41 48 13 [6]\\
$r_\mathrm{core}$ &  (pc) & 1.2$\pm$0.1 & 1.2$\pm$0.1 & 0.048$\pm$0.031\\
$r_4$ &  (pc) & 4.6$\pm$0.4& 4.6$\pm$0.4 & 0.19$\pm$0.12\\
$N_{\mathrm{core}}$ &  (stars) & 1400$\pm$100 & 1300$\pm$100 & 17$\pm$10\\
$N_{4}$ &  (stars) &  5600$\pm$250 & 5400$\pm$240 & 72$\pm$20 \\
$N_{\mathrm{FOV}}$ &  (stars) &  5700$\pm$250 & 5500$\pm$240 & 240$\pm$40 \\
$\Sigma_0$ &  (stars\,pc$^{-2}$) &470$\pm$80 & 460$\pm$60 & 3400: \\
$\rho_0$ &  (stars\,pc$^{-3}$) & 200$\pm$50& 200$\pm$50 & 35000: \\
$\epsilon$ &   & 0.34$\pm$0.05& 0.33$\pm$0.05 & 0.66$\pm$0.46\\
$\phi$ &  (degrees) & 159$\pm$4& 162$\pm$5 & 165$\pm$9\\
$\mathcal{L}$ & & 1599 & \multicolumn{2}{c}{1618}\\
BIC &   & $-$3158~~ &\multicolumn{2}{c}{$-$3194~~~} \\
AIC &   &  $-$3186~~ & \multicolumn{2}{c}{$-$3223~~~}\\
\enddata
\tablecomments{The best-fit parameters for the isothermal ellipsoids used to model the projected surface density. The results for a one-component model (single cluster) and a two-component model (main cluster $+$ subcluster) are shown. Rows~1-2: The coordinates of the ellipsoid centers,  with uncertainty given in brackets. Row~3: The harmonic-mean radius of the isodensity ellipse enclosing the cluster core. Row~4: A characteristic radius four times as large as the core radius. Rows~5-7: The number of stars assigned to each component within 1 core radius, within 4 core radii, and within the field of view. Rows~8-9: The surface density and the volume density at the center of the ellipsoids. Row~10: Ellipticity. Row~11: Ellipsoid position angle in degrees east from north. Rows~12-14: Log likelihood, BIC, and AIC of the model. Entries with units of "stars" have been corrected for sample incompleteness and represent the intrinsic stellar population down to 0.08~$M_\odot$. The reported uncertainties represent statistical uncertainty in both estimation of number of stars and model fitting, but exclude possible systematic error discussed in \S\ref{cluster_properties.sec}.  }
\end{deluxetable}

The total number of stars in a cluster described by an isothermal ellipsoid model depends on the outer radius of the cluster. However, the outer radius can be difficult to constrain because cluster members will be most thinly distributed near the outer edge of the cluster. A cluster like NGC~6231 subtends a large area on the sky, outside the fields of view of the various X-ray observations, so large surveys would be needed to determine the outer edge of the cluster. \citet{2015MNRAS.448.1687S} use the 2MASS catalog to estimate where the spatial over-density of stars meets the background level, and they report a radius of 36.2~pc. Our investigation in \S\ref{obaf} shows that these stars are not distributed isotropically around NGC~6231, being mostly distributed to the north and east of the main cluster.  With only spatial data, it is impossible to determine whether these stars are part of NGC~6231 or part of other clusters or associations in Sco~OB1.

There are several methods to describe the number of stars in NGC~6231 in a way that can be compared to observations of other clusters. The number of stars projected in the {\it Chandra} field of view is $N_{\mathrm{core}}\approx5700$~stars; the number of stars projected in an ellipse with characteristic radius $r_4$ is $N_{4}\approx5400$~stars; and the number of stars within a three-dimensional ellipsoid with characteristic radius $r_4$ is $\sim$4300~stars. These subtle geometric differences in definitions do not strongly affect results---the main point being that, in each case, a large number of cluster members may exist outside the region being considered.

The total mass of stars in the cluster depends on the binary fraction of its low-mass stars, which is not yet well constrained by observation. We follow \citet{2013MNRAS.429.1725M} to estimate the mean mass of single stars and multiple-star systems based on the \citet{2003PASP..115..763C} IMFs. Equation~25 from \citet{2013MNRAS.429.1725M} gives an average mass $\bar{m}=0.61~M_\odot$ for the mass range 0.08--150~$M_\odot$ and an average mass of $\bar{m}=0.78~M_\odot$ for systems. When this uncertainty is combined with the statistical uncertainty on total number of stars, the total mass of stars projected within the 4-core radius ellipse, down to the hydrogen-burning limit, is in the range 3300--4200~$M_\odot$. 

\section{Segregation of Stars by Stellar Mass \label{seg.sec}}

The spatial distribution of stars, with their masses indicated, is shown in Figure~\ref{mass_seg.fig} (left). From this figure it can be seen that stars of various masses are mixed together, with both high-mass stars and low-mass stars concentrated toward the center of the cluster. It appears that the O and B stars are relatively more likely to be found in the cluster center compared to low-mass stars. However, several high-mass stars are also located outside the cluster center, including the O9.5~III star HD~152076 (central distance of 13$^\prime$) and the O9~III+O9.7~V system HD~152247 (central distance of 11$^\prime$).  Only stars above the mass completeness limit of 0.5~$M_\odot$ are shown, so as to avoid effects of insensitivity to lower mass stars, either from instrumental effects or from crowding of stars. A mass-complete sample is important for establishing the reality of mass segregation because incompleteness could masquerade as an erroneous signature of mass segregation \citep{2009A&A...495..147A}.

\subsection{Statistical Tests for Mass Segregation}

\begin{figure*}[t]
\centering
\includegraphics[width=0.50\textwidth]{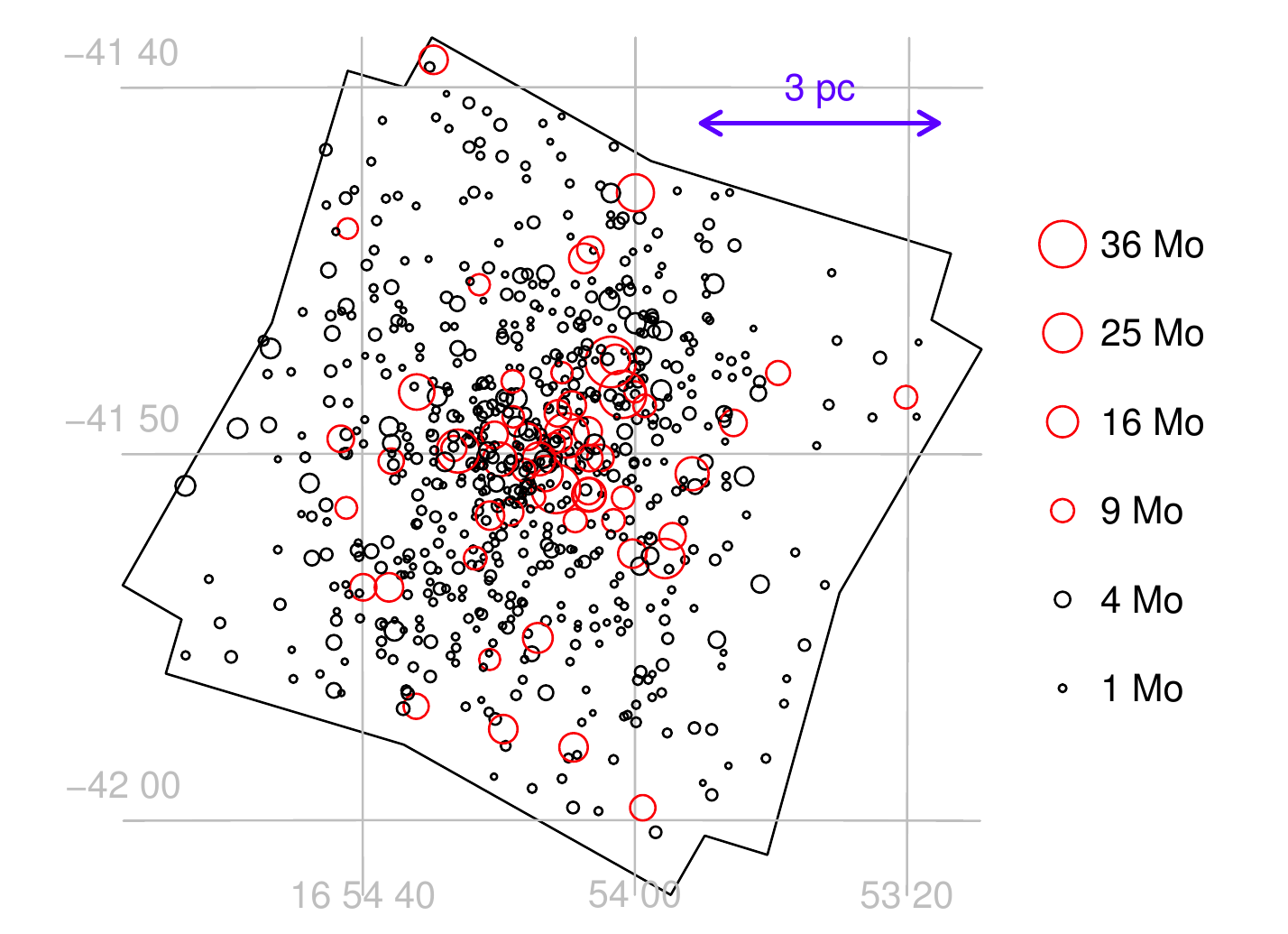} 
\includegraphics[width=0.48\textwidth]{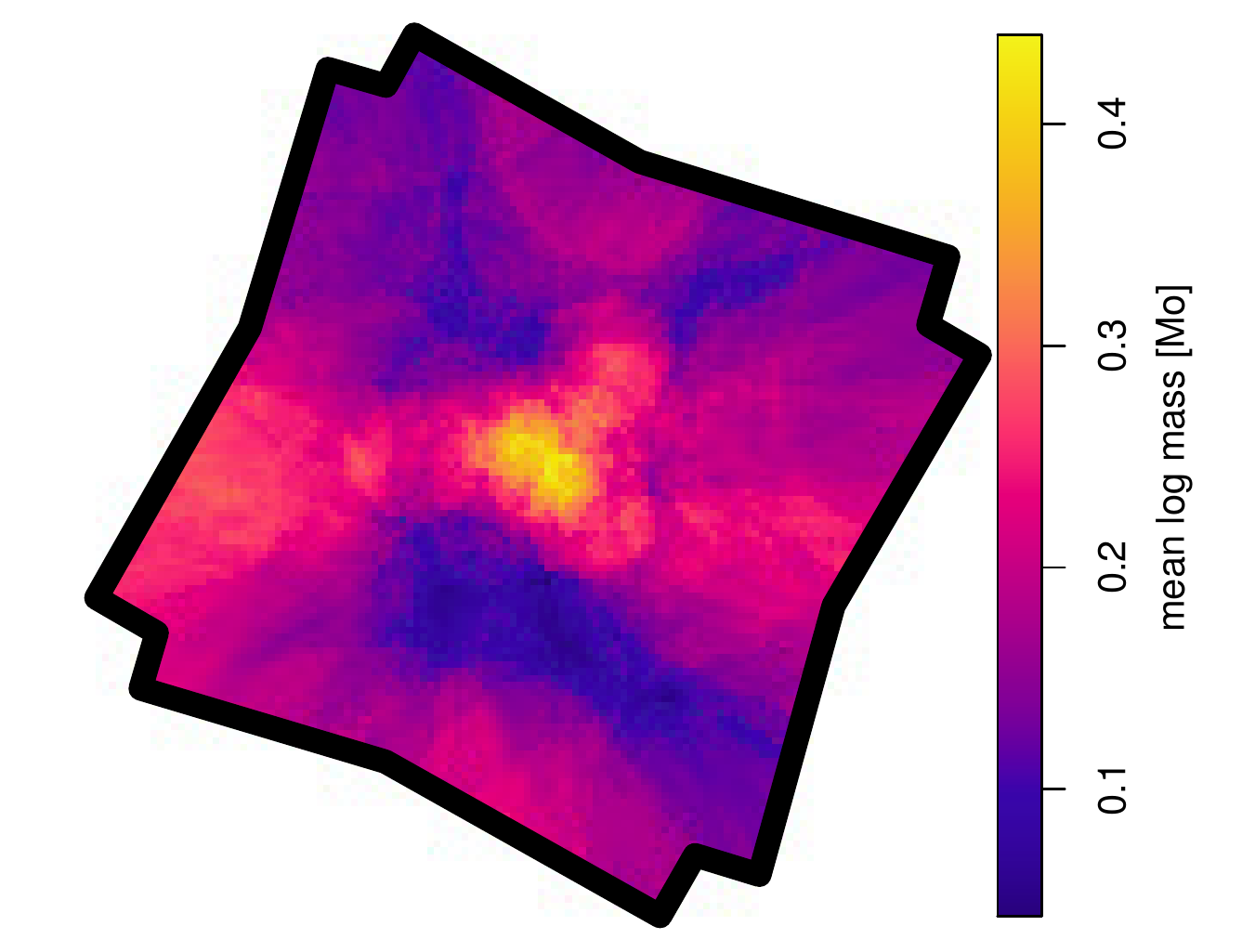} 
\caption{Left: Stars in flux-complete sample with mass estimates $>$0.5~$M_\odot$ are plotted on the {\it Chandra}/ACIS-I field of view for NGC~6231. The area of the circles is proportional to the estimated stellar masses, and stars with $M<7$\,$M_\odot$ are plotted with black circles and stars with $M>7$\,$M_\odot$ are plotted with red circles. Right: Adaptively smoothed mean values of $\log M$. The smoothing uses a Gaussian kernel, with a width $\sigma$ equal to the distance to the 10th nearest point. The color scale shows the mean $\log M$ values over a range from $\sim$1--5\,$M_\odot$.
 \label{mass_seg.fig}}
\end{figure*}

Figure~\ref{mass_seg.fig} (right) shows an estimate of the expected value of $\log M$ as a function of position of a star in the field of view. Interpolations of properties of points to generate a map of expected values is a well-known method of statistical analysis \citep[e.g.,][]{olea2000geostatistics}.
Due to large differences in surface density, we perform adaptive kernel smoothing, where the kernel bandwidth at a point $(x,y)$ is set to the distance to the 100th nearest star. The resulting map shows the highest average $\log M$ at the center of the cluster, as would be expected if the cluster were mass segregated. 
The mean value of this peak is $\langle \log M/M_\odot \rangle = 0.44$ ($\sim$2.8~$M_\odot$). This peak value is relatively low because of the large population of low-mass stars relative to high-mass stars, even near the cluster center. 

The statistical significance of the peak in the map can be judged through Monte Carlo simulations. For these simulations, stellar masses are randomly permuted among the stars to simulate a case in which stellar mass is independent of position, and a surface density map is generated in the same way. One thousand simulations are performed, and the maximum peak in the simulated maps is recorded. Based on the distribution of simulated peak values, the observed value peak of $\langle log M/M_\odot \rangle = 0.44$ has a $p$-value of $<$0.01.  

\begin{figure}[t]
\centering
\includegraphics[width=0.40\textwidth]{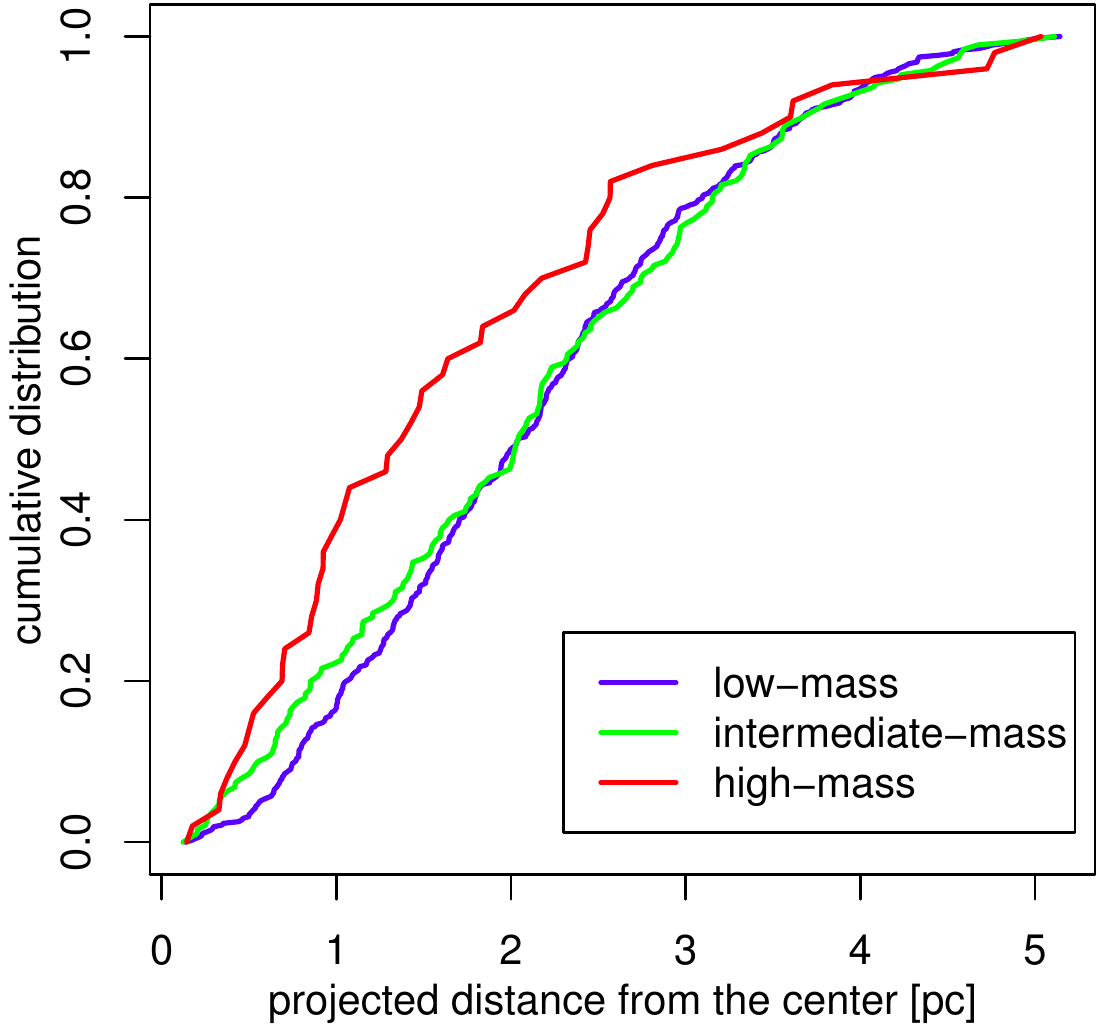} 
\caption{Cumulative distributions of projected distance from the center of the cluster for three different mass strata, which are divided at 1.8\,$M_\odot$ and 7.9\,$M_\odot$ into low- (blue), intermediate- (green), and high-mass (red) strata. The Anderson-Darling two-sample test gives $p$-values of 0.23 (low vs.\ intermediate), 0.001 (low vs.\ high) and 0.03 (intermediate vs.\ high). 
 \label{mass_seg2.fig}}
\end{figure}

Comparison of radial-distance distributions of stars in different mass strata (or comparison of stellar mass distributions in radial bins) has been a common method of testing for mass segregation \citep{1998ApJ...492..540H,2002MNRAS.331..245D,2006AJ....132..253S,2010ApJ...725.2485K}. Figure~\ref{mass_seg2.fig} shows cumulative distributions for three mass strata, the low-mass stars ($\log M/M_\odot < 0.25$), intermediate-mass stars ($0.25 < \log M/M_\odot <0.9$), and high-mass stars ($ \log M/M_\odot >0.9$).  The radial distributions of stars in these groups are compared using the two-sample Anderson-Darling test \citep{stephens1974edf}. Only the stratum containing the high-mass stars shows any difference in radial distribution (with $p$-value of 0.001 and 0.03 when compared to low- and intermediate-mass strata, respectively), while the low- and intermediate-mass strata have radial distributions that are very similar to each other. This finding agrees with observations of NGC~6231 by \citet{1998A&A...333..897R} that only the most massive stars are segregated, but intermediate stars are well mixed. In contrast, the study of mass segregation in 17 MYStIX star-forming regions by Kuhn et al.\ (in preparation) reveals many cases where even low-mass stars do appear to be strongly segregated by mass; but this is not the case for NGC~6231. 

\subsection{An Empirical Model of Mass Segregation}

\begin{figure}
\centering
\includegraphics[width=0.45\textwidth]{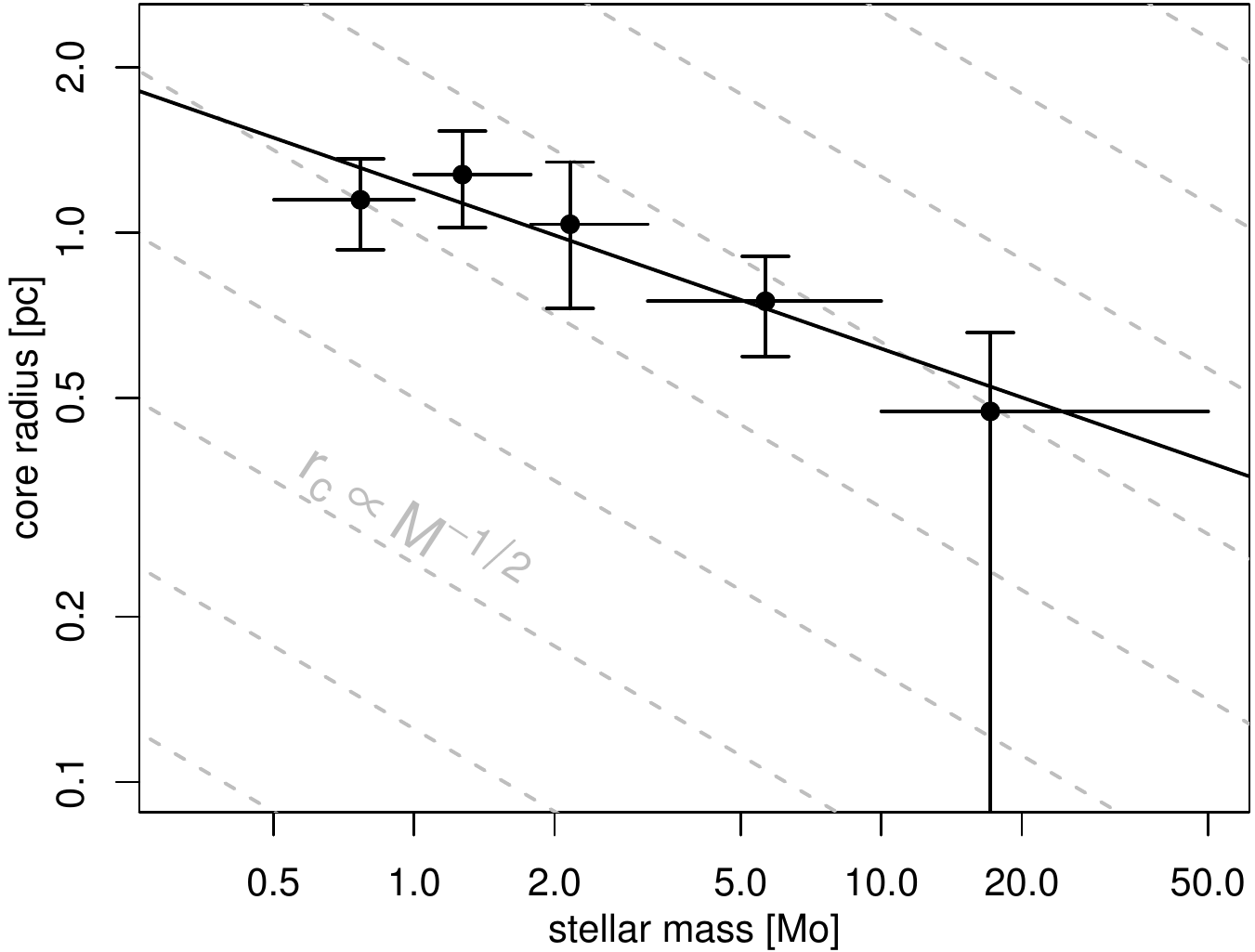} 
\caption{The best-fit model-cluster radius is plotted versus stellar mass, based on model fits to 5 sets of stars stratified by stellar mass. The horizontal lines show the range of stellar mass included in each sample, which are divided at $10^{-0.5}$, 1, $10^{0.25}$, $10^{0.5}$, and 10\,$M_\odot$, and the position of the point is the mean mass in each sample. The error bars on the core radius are based on the Hessian matrix at the maximum of the likelihood function,  which appear asymmetric on the logarithmic axis scale. Gray dashed lines indicate possible radius--mass relations for a cluster/association with energy equipartition. The solid black line shows an empirical ordinary-least-squares regression to the data. 
 \label{equipartition.fig}}
\end{figure}

To further investigate the effect of stellar masses on the distributions of stars, we subdivided the sample by stellar mass and fit the subsamples with the single isothermal-ellipsoid model from Section~\ref{mixture.sec}. (The small subcluster to the northwest is ignored here.) Five samples were used, divided at $\log M/M_\odot = 0.0, 0.25, 0.5, 1$, containing 251, 223, 126, 75, and 41 stars, respectively.  Figure~\ref{equipartition.fig} shows the plot of subcluster core radius vs.\ stellar mass. The points have an abscissa value equal to the mean mass of stars in a subsample (the horizontal bars show the range of masses in the subsample) and an ordinate value equal to the core radius (the vertical error bars show the $1\sigma$ uncertainty on core radius). The gray, dashed lines show relations of $r_c \propto M^{-1/2}$ for comparison. 

Figure~\ref{equipartition.fig} shows that, aside from the second mass bin ($0.5<M<1.0$~$M_\odot$), the core radius decreases monotonically with increasing stellar mass. However, for the range 0.5--3~$M_\odot$, the statistical uncertainties on core radius show that core radius is not statistically different for the first three mass bins. Although there is a persistent decrease between 1~$M_\odot$ and 50~$M_\odot$, the difference in core radius only becomes statistically significant for the highest mass bin (stars with $M>10$~$M_\odot$), which agrees with the results of the Anderson-Darling tests which only show mass segregation for high mass stars. 
The relation between core radius and mean stellar mass is described by an empirical relation $r_c\propto M^{-0.29\pm0.06}$ (black line) found using weighted ordinary least-squares regression using the bins shown, where weights are equal to the reciprocal of the estimated uncertainty.  Gray dashed lines with a $\sigma_v(m)\propto m^{-1/2}$ relation are shown for comparison.

\section{Age Distribution\label{age.sec}}

In \citetalias{KMG17} stellar ages are estimated using two independent techniques. The first estimates (denoted $Age_{JX}$) are based on X-ray and near-infrared photometry, using the method from \citet{2014ApJ...787..109G,2014ApJ...787..108G}.  The second estimates (denoted $Age_{VI}$) are based on the optical $V$ vs.\ $V-I$ color-magnitude diagram. Age estimates from both methods are calibrated using the \citet{2000A&A...358..593S} models. In addition, the $Age_{JX}$ method was based on relations derived for pre-main-sequence stars with ages $<$5~Myr old, so stars with ages greater than 5~Myr will be assigned 5~Myr as a lower limit. As reported by \citetalias{KMG17}, the median $Age_{JX}$ for stars in NGC~6231 is 3.2~Myr, which is very similar to the median $Age_{VI}$ value of 3.3~Myr. While statistical uncertainties on ages of individual stars may be large, statistical uncertainties on the median ages of sufficiently large sample of stars will be smaller. Thus, median-age estimates can be used to identify spatial age gradients.

In many young star-forming regions, differences in ages of groups of stars are of the order $\sim$1~Myr \citep{2014ApJ...787..108G}. This supports a model of star formation in which all stars do not form at in a monolithic cluster in a single cluster-crossing timescale. Instead, stars form either over multiple free-fall timescales or in multiple, independent subclusters that form at different times. In the two cases from the MYStIX study with sufficient data quality, the Orion Nebula Cluster and NGC~2024, stars within an individual cluster also showed a radial gradient in stellar age, with the youngest stars nearest the cluster center and the older stars in the cluster periphery \citep{2014ApJ...787..109G}. This was regarded as an unexpected result because stars are typically expected to form first where gas in molecular clouds is densest, which would be in the centers of clusters. Additional cases are reported by K.\ V.\ Getman et al. (2017, in preparation) indicating that age gradients are common in young clusters.

\begin{figure*}
\centering
\includegraphics[width=0.45\textwidth]{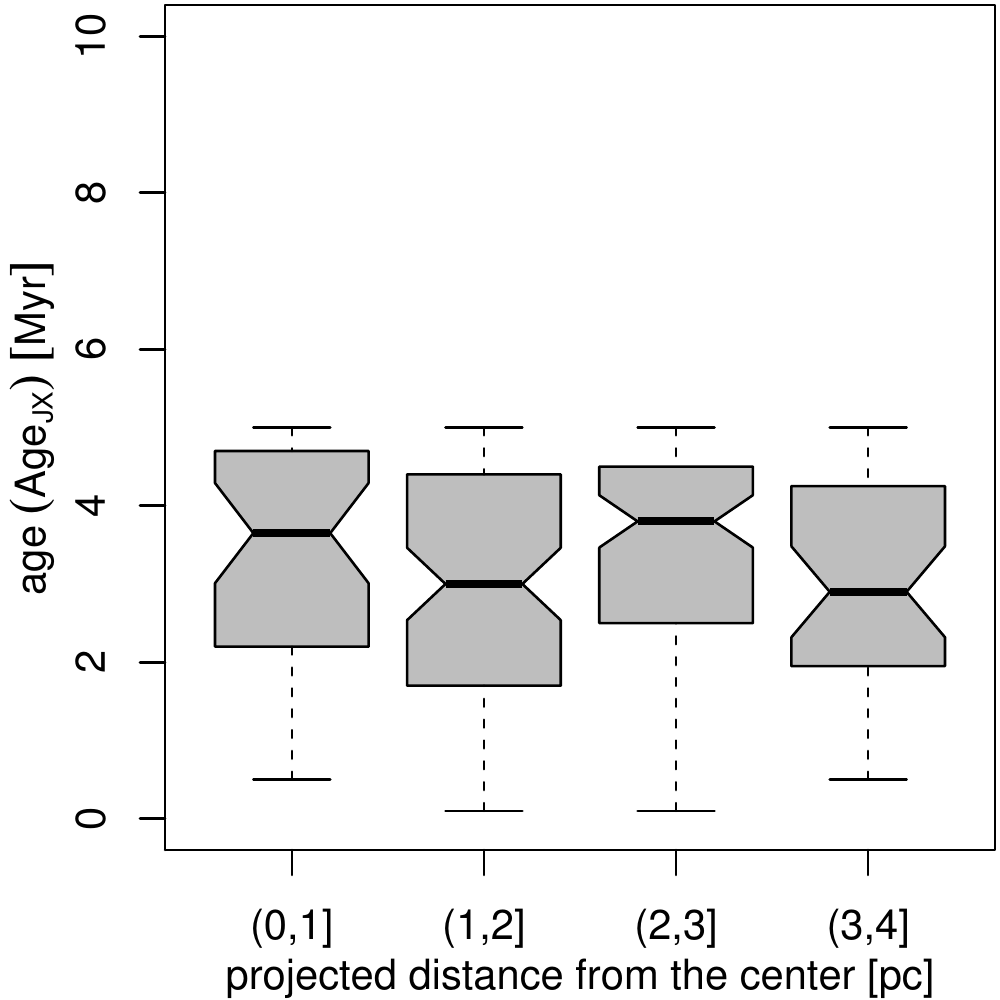} 
\includegraphics[width=0.45\textwidth]{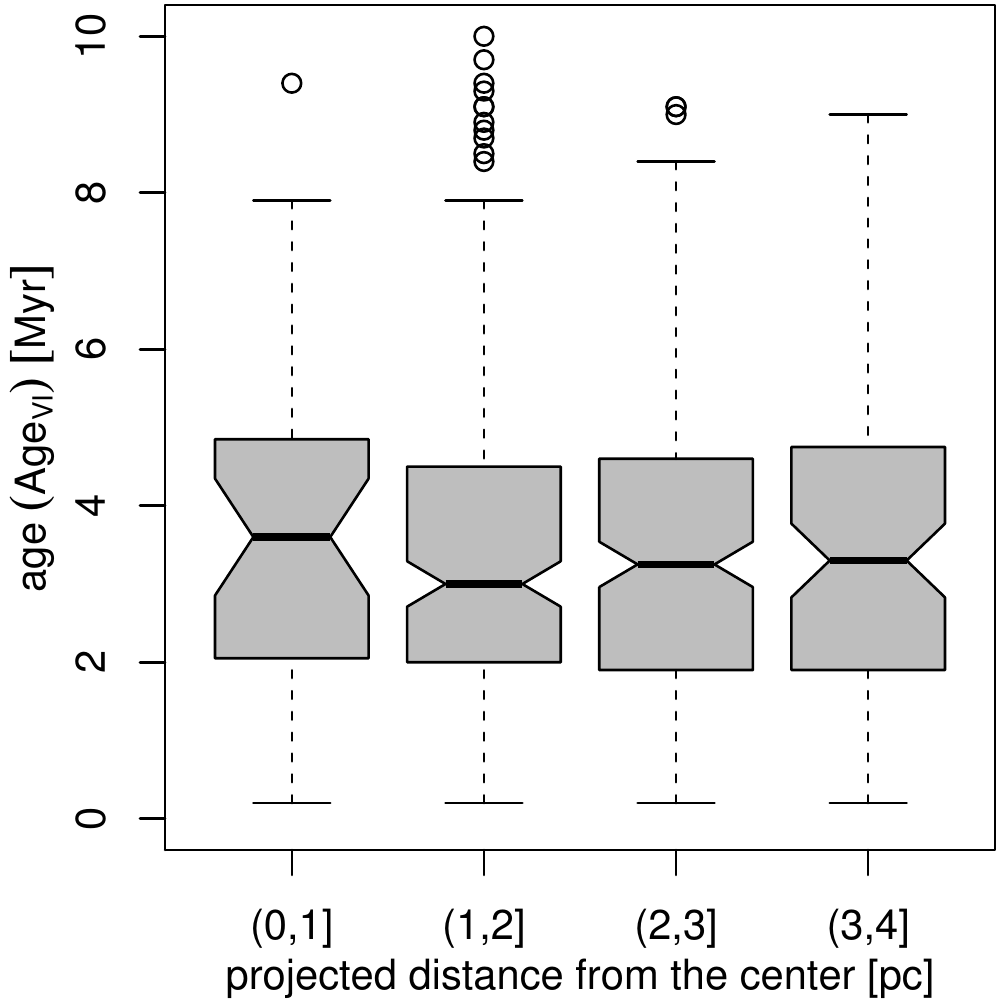} 
\caption{Box-And-Whisker plots of the $Age_{JX}$ (left) and $Age_{VI}$ (right) distributions for stars in several radial bins. The boxes indicate the 25\%, 50\% (median), and 75\% quartiles for the Age$_{JX}$ values for stars 0--1\,pc, 1--2\,pc, 2--3\,pc, and 3--4\,pc from the cluster center (from the single-ellipsoid model). Notches indicate uncertainty on the sample median. The whiskers are indicate the range of the data, up to 1.5 times the interquartile ratio, and outliers are drawn as open circles. The data on these plots span different ranges because $Age_{JX}$ is only estimated for stars with ages $<$5~Myr and stars that appear older are assigned 5~$Myr$ as a lower limit. Nevertheless, both plots show very similar median ages. 
\label{baw.fig}}
\end{figure*}

The NGC~6231 cluster appears to have an age spread of 2--7~Myr, noted by previous studies \citep{2007MNRAS.377..945S,2013AJ....145...37S,2016arXiv160708860D} and \citetalias{KMG17}. With a median age of $\sim$3.2~Myr, NGC~6231 is older than most MYStIX star-forming regions, including the Orion Nebula Cluster and NGC~2024, and star formation has ended in NGC~6231, so stars cannot be extremely young (e.g., $<$0.1~Myr). Figure~\ref{baw.fig} shows the median $Age_{JX}$ and $Age_{VI}$ values (25\%, 75\%, and uncertainties on the median are also shown) for stars at several projected distances from the cluster center. The range of ages (median, 1st-quartile, and 3rd-quartile) estimated using both independent techniques are very similar, although the $Age_{VI}$ ages have a greater range because they are not limited to 1--5~Myr. 

\begin{figure}
\centering
\includegraphics[width=0.5\textwidth]{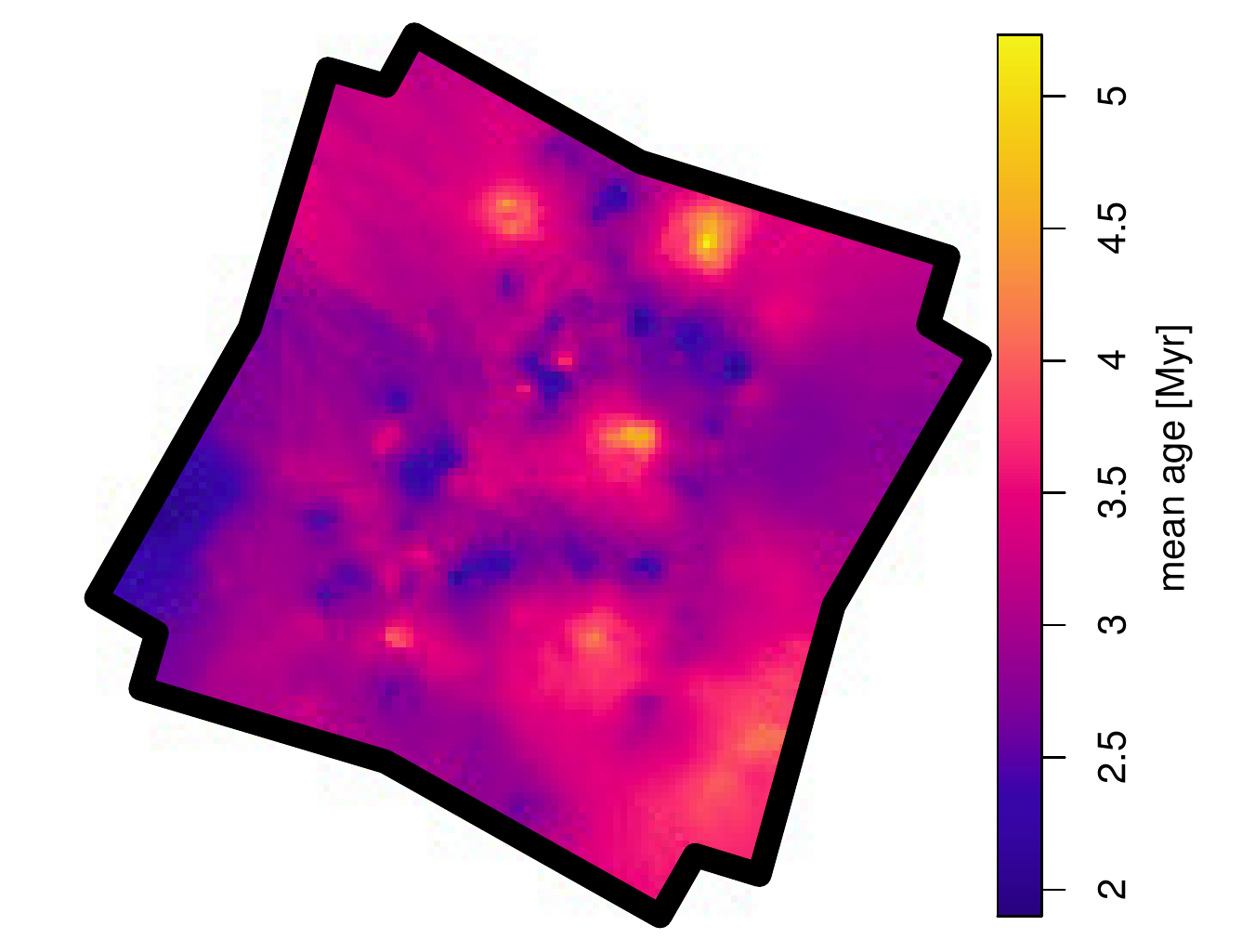} 
\caption{Adaptively smoothed mean ages from the Age$_{VI}$ method. The smoothing uses a Gaussian kernel, with a width $\sigma$ equal to the distance to the 5th nearest point. The color scale shows variations in mean calculated age from 2 to 5~Myr.
 \label{medianage.fig}}
\end{figure}

Median ages of the stellar population range from 3 to 4~Myr on a distance baseline ranging from 0 to 4~pc, and no systematic trend is seen. Thus, there is no large-scale radial age gradient. Figure~\ref{medianage.fig} shows an adaptively smoothed map of mean $Age_{VI}$. Here, variations in age throughout the field of view are small, and no global trend is evident. As before, Monte Carlo simulations can be used to determine if structures in a map may be the result of random fluctuations. For the map of mean ages, the low-amplitude structure is consistent with variations for a distribution where stellar age is independent of position. The lack of a radial age gradient implies that, either NGC~6231 never had a radial age gradient, or that a previously existing gradient disappeared with age due to dynamical mixing. This result is clearly different from the younger MYStIX clusters where a radial--age gradient has been observed.

\section{Larger-Scale Galactic Environment \label{largescale.sec}}

NGC~6231 extends beyond the {\it Chandra} field of view. However, in these regions, spatial distributions of stars can be mapped using the catalog of 295 near-infrared variables from VVV and the candidate early/mid-type stars from VPHAS+. Both these catalogs suffer from more incompleteness and more field-star contaminants than the {\it Chandra}-based catalog. Nevertheless, contaminants are expected to be distributed smoothly (except for possible patchy absorption) with some dependence on Galactic latitude, and candidate selection is not strongly affected by position. Thus, clustering of these sources is likely associated with real clusters or associations of stars. These two samples trace different types of populations. The amplitude of variability in the near-infrared decreases with a pre-main-sequence star's age \citep{2015AJ....150..132R}, so VVV variables will trace the youngest stellar population, while candidate early/mid-type stars will be less sensitive to age. 

\subsection{Modeling Clusters of Near-infrared Variables}

We use a mixture model approach to identify possible clusterings of variable stars measured in the VVV $K_s$ band. Given that there may be a high number of field stars, one component of the model will account for these objects. Field stars are expected to be smoothly distributed rather than clustered. However, the large field of view means that the projected density of field stars will vary with Galactic line of sight, mostly as a function of Galactic latitude $b$. The contribution of the unclustered field-star component in the mixture model is similar to the use of an unclustered component by \citet{2014ApJ...787..107K}, but here it is a model with several parameters. We use the flexible model form, 
\begin{equation}
\Sigma_\mathrm{unc.}(\ell,b) = C \exp[a_1 b + a_2 b^2],
\end{equation}
to describe variation in field-star densities, where the variable $C$ is the normalization of the model and the polynomial coefficients $a_1$ and $a_2$ are the parameters to be fit.

\begin{figure*}
\centering
\includegraphics[width=1.0\textwidth]{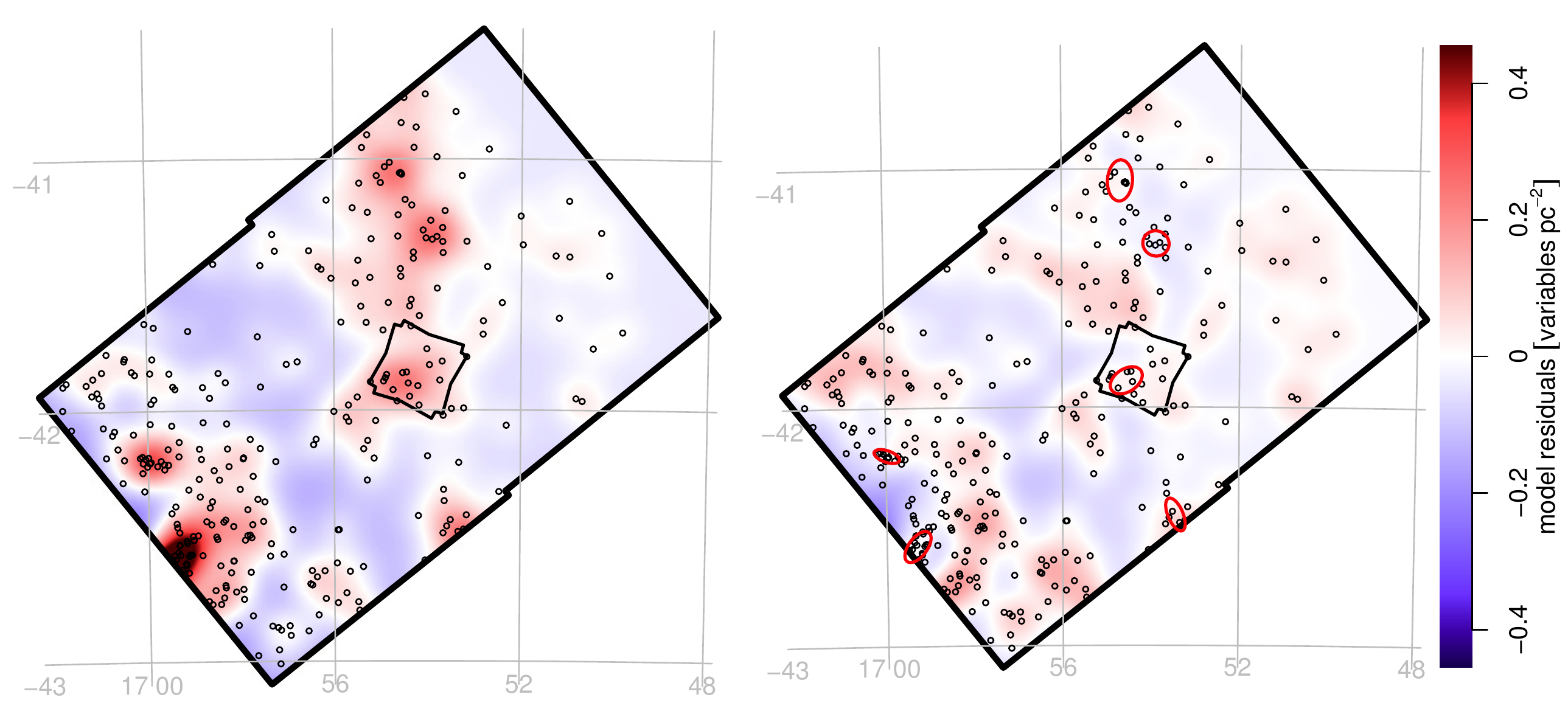} 
\caption{Spatial distribution of near-infrared variables in the VVV fields plotted on a residual map for two models: a smoothly varying model of Galactic field stars (left) and the multi-cluster mixture model (right). When the field stars are modeled, as shown in the left panel, possible clusters of stars stand out as positive (red) residuals. In the right panel, the locations of the clusters favored by the AIC are shown by red ellipses. The residuals at these locations are very small, implying a good fit to the data. The grid lines show right ascension and declination.
 \label{clustvar.fig}}
\end{figure*}

Figure~\ref{clustvar.fig} (left) shows the model residuals when the data have been fit with the unclustered model (the hypothesis that all variables are field stars). The best-fit parameters of the density gradient are $a_1=-1.47$~deg$^{-1}$ and $a_2=0.26$~deg$^{-2}$. Several density peaks are not well modeled, leading to high residuals (red patches). The residuals include a peak associated with NGC~6231, several peaks to the south-east of NGC~6231 near the Galactic plane, and several peaks to the north and south of NGC~6231. 

Next, we test various mixture models, composed of $G$ ``isothermal ellipsoid'' components and an ``unclustered'' component. The identification of multiple statistical clusters follows the same method as \citet{2014ApJ...787..107K}. Models with $G=0$, 1, 2, 3, 4, 5, and 6 are fit, and the AIC and BIC calculated. For this model, the number of parameters is $k=6G+3$, so the penalty for each additional component is 12 for the AIC and 34 for the BIC. The AIC values are 1767, 1771, 1717, 1668, 1671, 1668, and, 1666 and the BIC values are 1778, 1804, 1773, 1745, 1770, 1789, and 1810, for $G=0$, 1, 2, 3, 4, 5, and 6 clusters respectively. Thus, the BIC clearly favors a 3-cluster model, while the AIC is consistent with models with 3--6 clusters. Note that groups of stars are only clusters in a statistical sense, while the physical nature of these groups is mostly uncertain.
 
 Table~\ref{vvvclust.tab} provides the list of cluster candidates identified from the 6-cluster model. Cluster candidates that are included in both the best AIC model and best BIC model are listed first, followed by cluster candidates that only appear in the best AIC model. These clusters are listed in approximate order of significance. Uncertainties on model parameters related to cluster shape are large. For example, in all cases the radius of the cluster core is poorly constrained. Thus, we do not report cluster core radius, ellipticity, or orientation. The number of stars reported in the table is the total number of stars in the cluster. Figure~\ref{clustvar.fig} (right) shows the residual map for a 6-cluster model that is favored by the AIC. Red ellipses show the locations of the clusters. 
 
 \begin{deluxetable}{lrrcr}
\tablecaption{Clusters of VVV Variables \label{vvvclust.tab}}
\tabletypesize{\small}\tablewidth{0pt}
\tablehead{
\colhead{No} & \colhead{RA} &\colhead{Dec} & \colhead{Unc.} & \colhead{$N_\star$}\\
\colhead{} & \colhead{(J2000)} &\colhead{(J2000)} & \colhead{(arcmin)} & \colhead{(stars)}
}
\startdata
\multicolumn{4}{l}{Favored by the AIC and BIC}\\
1 & 16 59 18& $-$42 34 50& [1.7 1.9]& 53~~~~~\\
2 & 16 59 58& $-$42 12 00& [0.9 0.6]& 21~~~~~\\
3 & 16 54 28& $-$41 02 50& [1.4 2.0]& 30~~~~~\\
\multicolumn{4}{l}{Consistent with the AIC} \\
4 & 16 54 33& $-$41 53 20& [2.1 1.6]& 33~~~~~\\
5 & 16 53 53& $-$41 18 50& [1.2 0.8]& 22~~~~~\\
6 & 16 53 26& $-$42 27 00& [1.2 2.0]& 11~~~~~\\
\enddata
\tablecomments{Clusters of VVV $K_s$-band variables identified using the mixture model analysis. The cluster supported by both AIC and BIC is listed first, followed by clusters supported by only the AIC. Column~1:  Cluster number. Columns~2--4: Celestial coordinates of cluster center, and uncertainty on these positions in arcminutes. Column~5: Number of variable stars in the cluster (integrated over the entire field of view). Model properties such as core-radius, ellipticity, and orientation are poorly constrained, so we do not report these value. Note that cluster \#4 is NGC~6231. }
\end{deluxetable}

\subsection{Relation of Clusters of Near-infrared Variables to NGC~6231}

VVV tiles ``d148'' and ``d110'' cover large angular areas in the Galactic plane, so it is likely that multiple, unrelated clusters of young stars will be identified within the field of view. The densest cluster \#1 of VVV variables at coordinates 16\hr59\mn18\se $-$42\de34\am50\as\ is spatially coincident with the cluster DBSB~176 associated with IRAS~16558-4228 \citep{2003A&A...400..533D,2007ApJ...659.1360W}.  Mid-infrared images of DBSB~176 from the GLIMPSE survey \citep{2003PASP..115..953B} show significant nebulosity in the region, suggesting active star formation. This cluster contains more VVV variables than NGC~6231, which may be explained if it is younger than NGC~6231. To the north east of DBSB~176 is another strong over-density of VVV variables at coordinates 16\hr59\mn58\se$-$42\de12\am00\as.

A cluster of VVV variables is associated with the center of NGC~6231, but this cluster is found in the best AIC model, but not the best BIC model. The relatively low number of VVV variables in NGC~6231 may be related to the relatively low disk fraction in NGC~6231 because Class~III pre-main-sequence stars typically have lower near-infrared variability amplitudes than Class~0/I/II young stellar objects. Although the analysis by \citet{2015MNRAS.448.1687S} suggested a radius for NGC~6231 of 68~arcmin, in the spatial distribution of VVV variables, there is no radially symmetric over-density of this size. However, the VVV variables include only a small fraction of the cluster members, and may not be sufficient for probing low-density distributions of stars that define cluster's outer boundary.

Rather than a radially symmetric distribution of VVV variables around NGC~6231, there are clumps with higher densities of variables to the north of NGC~6231, modeled by candidates at 16\hr54\mn28\se$-$41\de02\am50\as\ and 16\hr53\mn53\se$-$41\de18\am50\as. These two groups of stars lie between NGC~6231 and Tr~24. If they do lie at the same distance of the rest of this complex, they could indicate further subclustered structure in the outer northern portions of NGC~6231. However, no information is currently available about their distance.

The least statistically significant cluster candidate in the 6-cluster model is located at 16\hr53\mn26\se$-$42\de27\am00\as. This is near enough to NGC~6231 that it is plausible that it is related to the Sco~OB1 association, but it may also be an unrelated cluster or association in the Galactic plane. 

~

\subsection{Distributions of Candidate Early- and Intermediate-type Stars in VPHAS+\label{obaf}}

\begin{figure}
\centering
\includegraphics[width=0.45\textwidth]{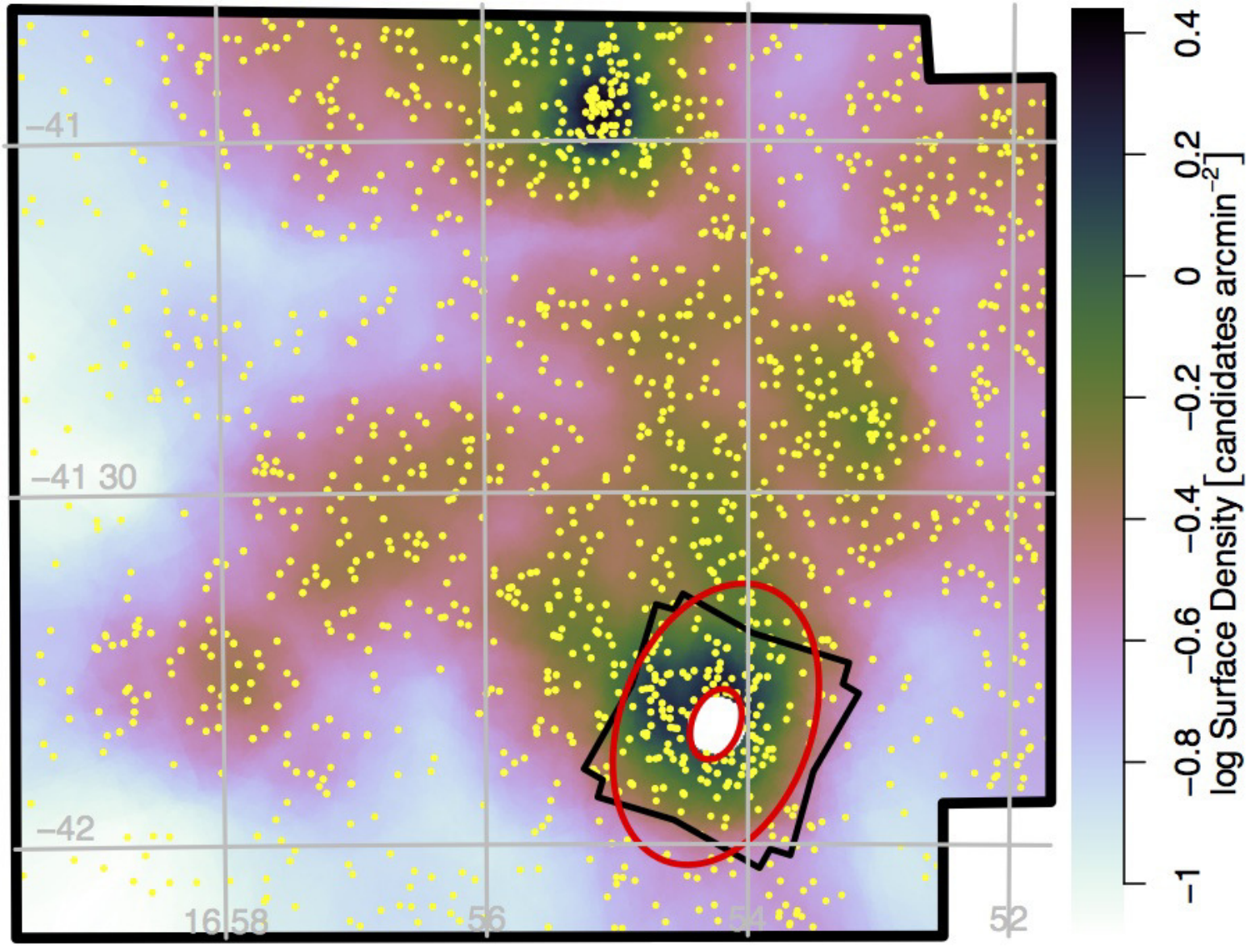} 
\caption{Spatial distribution of candidate O, B, A, and F-type stars in the VPHAS+ fields around NGC~6231. Yellow points mark the candidates selected from the $g$ vs.\ $g-i$ diagram. The adaptively smoothed surface-density map is shown by the color map. The {\it Chandra} field of view is shown as a black polygon, and the 1-core radius and the 4-core radii ellipses for the main NGC~6231 cluster are shown in red. Selection of stars is impeded in the center of NGC~6231 due to bright stars, so this area is not included in the analysis. 
 \label{vphas.fig}}
\end{figure}

The spatial distribution of the candidate O, B, A, and F-type stars from the VPHAS+ survey is shown in Figure~\ref{vphas.fig}. Surface density varies by 1.4~dex. We exclude a region around the center of NGC~6231 from the diagram, because the wings of bright O-stars in NGC~6231 inhibits the detection of A and F stars there. 

These stars are also not distributed around NGC~6231 evenly, but concentrated to the north of the cluster instead. The two main peaks in surface density correspond to NGC~6231 and another cluster at coordinates 16\hr55\mn10\se$-$39\de57\am00\as, just to the north east of VVV variable cluster \#3 from Table~\ref{vvvclust.tab}.

The core region of NGC~6231 derived in \S\ref{mixture.sec} is shown as a red ellipse on the map in Figure~\ref{vphas.fig}, and an additional ellipse 4 times the size of the core is also shown. It is difficult to define an outer boundary to the cluster because the over-density associated with NGC~6231 blends into the over-densities associated with the larger-scale Sco~OB1 region. Without measurements of stellar kinematics or three-dimensional coordinates of stars in the Sco~OB1 region, it is impossible to determine whether there is a meaningful astrophysical distinction between the NGC~6231 stellar population and the Sco~OB1 population. 

The relation between NGC~6231 members and Sco~OB1 members may resemble the relation between members of the Orion Nebula Cluster and members of the Orion Molecular cloud. In both cases the distribution of stars resembles a smooth, centrally concentrated cluster on a smaller scale, and a more elongated and clumpy distribution at on a larger scale \citep{1998ApJ...492..540H,2012AJ....144..192M,2016AJ....151....5M}.

\section{Discussion:\\ Cluster Formation and Fate \label{discussion.sec}}

\subsection{Summary of Observational Results}\label{summary.sec}

The main cluster properties, as listed below, can be used to place the cluster in a Galactic context and to test theoretical models for cluster formation and evolution.  
\begin{enumerate}
\item The median age of the cluster is estimated to be $\sim$3.2~Myr, but there may be systematic uncertainty in age. There is evidence of a large age spread with stellar age estimates ranging from 2 to 7~Myr.

\item The density of stars at the center of the cluster is $\rho_0 = 200\pm50$~stars~pc$^{-3}$ (or a column density of $\Sigma_0=460\pm60$~stars~pc$^{-2}$. 

\item The total number of stars in the cluster has a lower limit of 5700$\pm$250~stars, down to the hydrogen burning mass limit. However, the ability to estimate total cluster population is limited by the {\it Chandra} field of view. This population corresponds to a total stellar mass of 3300--4200~$M_\odot$.

\item Gas mass does not contribute significantly to the cluster's gravitational potential.

\item The radial density distribution of stars resembles an isothermal-ellipsoid distribution with significant ($\epsilon=0.33\pm0.05$) elongation of the cluster. The measured isothermal-ellipsoid core radius is 1.2$\pm$0.1~pc.

\item There is a second mode in the surface density map, which corresponds to a minor subcluster of stars with 4\% the population of the main cluster. 

\item No radial stellar-age stratification is evident. 

\item The spatial distribution of O and B stars shows statistically significant mass segregation, but lower-mass stars shown no sign of mass segregation. The dependence of spatial dispersion on stellar mass is described by a power-law relation $r_c\propto M^{-0.29}$.

\item The cluster follows the ``isothermal ellipsoid'' surface-density model out to at least 4 times the core radius. However, the distribution of pre-main-sequence stars in a several-square-degree field of view around the cluster is clumpy, rather than radially symmetric. 

\end{enumerate}

\subsection{Origin of NGC~6231}\label{origin.sec}

The initial properties of the molecular cloud and the early dynamical interactions of groups of stars may determine what type of star cluster or association is produced and whether it will survive as an open cluster \citep{2012MNRAS.426.3008K}. The lack of remaining cloud material and the dynamically evolved state of  NGC~6231 mean that formation characteristics such as the star-formation efficiency cannot be directly measured. However, some properties of the progenitor cloud can be inferred from the observed star cluster. 

\subsubsection{Filamentary Natal Cloud}
Young star clusters are often associated with filamentary molecular clouds on many size scales \citep[e.g.,][]{2010ApJ...719L.185J,2013A&A...554A..55H}. These elongated clouds may imprint their structure on the star clusters that form within them. For example, in the MYStIX study, several examples of ``linear chains of subclusters'' are noted \citep[][their Figure~5]{2014ApJ...787..107K}. In some of these, like DR~21, the young stars are deeply embedded in a massive infrared-dark filament. In NGC~1893, multiple subclusters are arranged linearly within a bubble, evacuated of molecular gas.

The distribution of VVV variables around NGC~6231 is not isotropic, but instead concentrated to the north of the cluster in two subclusters. This suggests that there is a population of young stars bridging the gap between NGC~6231 and Tr~24. The concentration of high and intermediate mass stars north of NGC~6231 also shows excess stars to the north of NGC~6231. This spatial distribution suggests that the progenitor cloud for NGC~6231 likely was filamentary with a north-south orientation.

An elongated cloud may also impart ellipticity on the clusters that form. For example, the Orion Nebula Cluster is elongated in a north-south direction \citep{1998ApJ...492..540H} approximately matching the orientation of the molecular filament. Simulations of the collapse of an elongated molecular cloud, starting with the likely initial state of Orion A cloud \citep{2007ApJ...654..988H}, do produce an elongated young star cluster \citep{2015ApJ...815...27K}.

The core of NGC~6231 is also clearly elongated with statistically significant ellipticity ($\epsilon=0.33\pm0.05$). This is similar to the ellipticity of the Orion Nebula Cluster ($\epsilon=0.3$--0.5) measured by \citet{2014ApJ...787..107K} using the same method. The orientation of the core is close to being north-south, with a moderate offset of $\sim$20$^\circ$. Thus, the elongation of the cluster could  be explained by cluster formation in a collapsing, elongated cloud, similar to the scenario described by \citet{2007ApJ...654..988H}.

\subsubsection{Multiplicity of Star-Forming Cloud Clumps}
Molecular clouds are typically clumpy, with fractal like density structures \citep{1998A&A...336..697S}. These can give rise to subclusters of stars \citep{2000ApJ...530..277E}, which often lie at the locations of dense molecular clumps \citep{2013ApJ...769..140Y,2009AJ....138..227F}. The clumpiness of the distribution of stars can vary from region to region, with subclusters of stars formed in individual cloud clumps possibly merging to form more centrally concentrated young star clusters \citep{2009MNRAS.397..954F}. \citet{2014ApJ...787..107K} find between 1 and 20 subclusters in each of the star-forming regions surveyed by MYStIX. 

Our study of NGC~6231 reveals a main cluster accounting for the vast majority of the stars seen in the {\it Chandra} field of view. However, a group of stars that is more densely clustered is identified as a statistically-significant subcluster offset from the cluster center. This group of stars most likely formed as a separate density enhancement in the molecular cloud, revealing that the initial cloud was clumpy. The distribution of VVV variables outside the {\it Chandra} field of view is also not uniform, suggesting that the progenitor cloud was clumpy on a larger spatial scale as well.

\subsubsection{Cluster Assembly}\label{assembly.sec}

Overall, the structure of NGC~6231 looks quite different from regions with ongoing star formation in the MYStIX study. The regions investigated by MYStIX show significant diversity in the spatial distributions of their stars -- some MYStIX rich clusters are smooth with ``simple'' structures, while others are ``clumpy'' and/or linear ``chains'' of subclusters \citep{2014ApJ...787..107K}. On one hand, NGC 6231 is very different from clusters like Orion Nebula Cluster and RCW~38 that are much smaller and more concentrated. On the other hand, it is also quite different from clusters like NGC~1893, NGC~6334, and Carina that have complicated clumpy or filamentary morphologies. NGC 6231 is most similar to NGC~2244 in Rosette -- both have similar ages, sizes, and numbers of stars, and both are located within evacuated bubbles.

Here we compare the physical properties of NGC~6231 (age, radius, number of stars, density) to the properties of the MYStIX clusters and subclusters. 
The diversity of these complexes may make it seem as if the (sub)clusters of stars they contain would not be directly comparable. Nevertheless, \citet{2015ApJ...812..131K} has shown that properties of (sub)clusters of stars in MYStIX, even in regions with different global morphology, do follow common trends. A comparison of NGC~6231 to subclusters in MYStIX may make sense if the individual subclusters in a clumpy distribution of stars are the building blocks for more massive clusters, and thus could yield insight into how NGC~6231 formed. In the following discussion, we highlight several examples from MYStIX of more fully-formed clusters that may be better analogs to NGC~6231. These include:  RCW~38 (Subcl.~B), NGC~2024, W40,  Pismis~24 (Subcl.~A in NGC~6357), G353.2+0.7 (Subcl.~F in NGC~6357), NGC~2362, NGC~6611 (Subcl.~B in the Eagle Nebula), and NGC~2244 (Subcl.~E in the Rosette Nebula). 

\begin{figure*}
\centering
\includegraphics[width=1.0\textwidth]{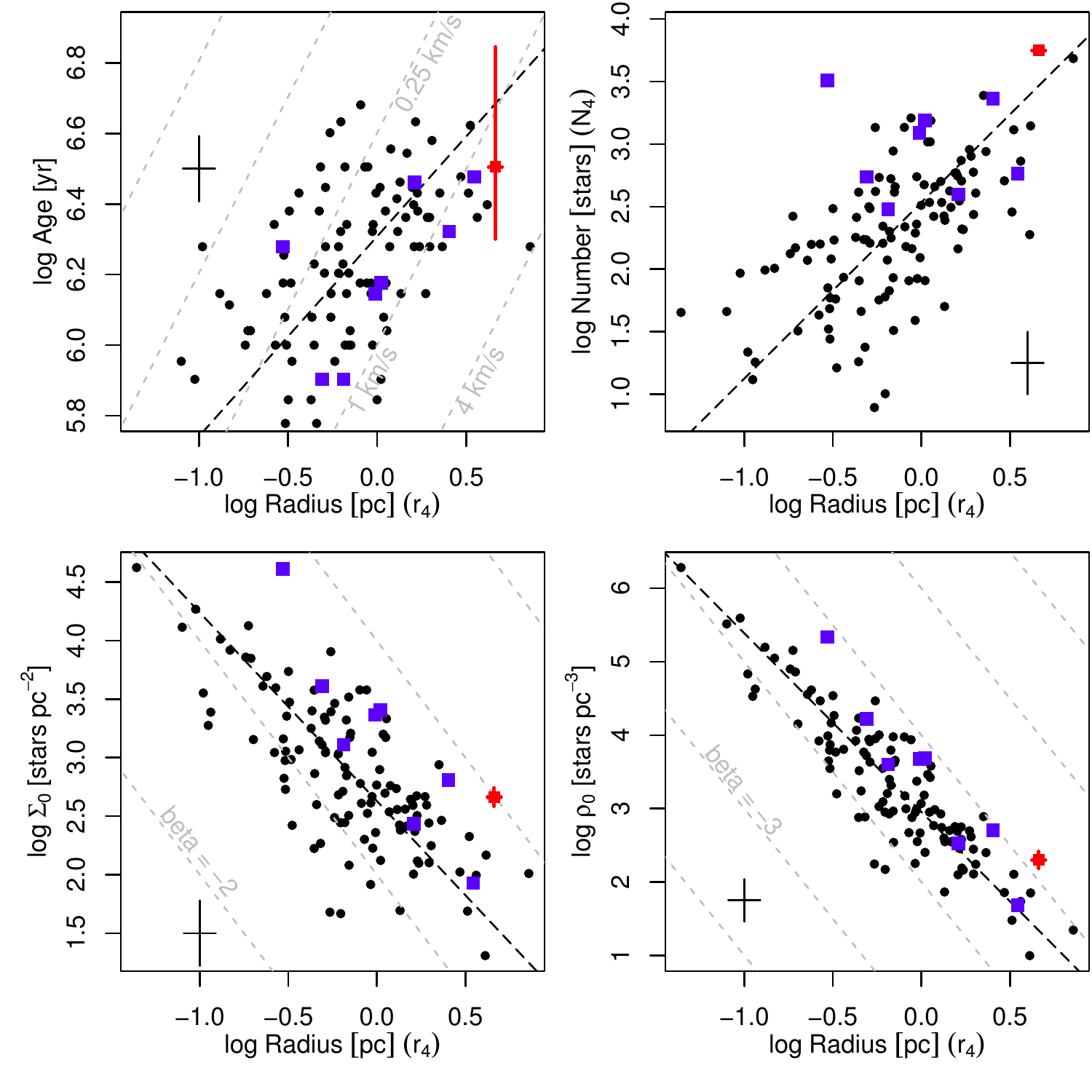} 
\caption{Cluster properties (radius, number of stars, central surface density, central volume density, and age) for NGC~6231 shown in comparison to the properties of subclusters of stars in other star-forming region from the MYStIX study. The red points mark NGC~6231 on these diagrams, and the uncertainties are shown by the red error bars. The MYStIX (sub)clusters are shown as black or blue points, typical uncertainties are indicated by black crosses, and  the orthogonal regression lines are shown as a dashed, black line. Gray lines indicate simplistic evolutionary tracks for clusters, whereby a cluster expands at a constant radial velocity and neither gains nor looses stars. For the MYStIX subclusters, only points with no missing values for age, $r_4$, $\Sigma_0$, and $\rho_0$ are included. Eight rich clusters in MYStIX, which make the best comparisons to NGC~6231, are highlighted in blue: RCW~38, NGC~2024, W40,  Pismis~24, G353.2+0.7, NGC~2362, NGC~6611, and NGC~2244 (from smallest to largest in radius).
\label{mystix_comparison.fig}}
\end{figure*}

Figure~\ref{mystix_comparison.fig} shows the cluster properties of the main NGC~6231 cluster (marked by a red square) compared to properties of MYStIX subclusters (black circles) and the highlighted MYStIX clusters (blue squares). The properties include cluster age, cluster core radius, number of stars (within 4 core radii), the projected central stellar density, and the central volumetric stellar density. The relations between these properties were investigated by \citet{2015ApJ...812..131K}, who show that this set of subcluster properties is statistically correlated, exhibiting a positive age--radius relation, a positive radius--number of stars relation, and a negative density--radius relation. The uncertainties on the properties of NGC~6231, which result from model fitting, from estimation of completeness, and from age estimation, are shown by the error bars. The regression lines for the relations found in MYStIX are also drawn.

NGC~6231's properties lie near the regression lines, at the older, larger, more-massive, and less-dense end of the distribution (Figure~\ref{mystix_comparison.fig}). NGC~2244 also lies at the same end of the parameter distributions. In the scatter plots, the point corresponding to NGC~2244 (Rosette Subcluster~E) is the blue point with the largest value of $r_4$. This cluster is very close in age and radius to NGC~6231; it is slightly less massive and less dense, but is still one of the most massive and least dense clusters in the MYStIX study. In contrast, most of the other rich clusters are also located near the regression lines, but with range of properties. The only massive cluster to significantly deviate from these trends is RCW~38, the blue point with the smallest value of $r_4$. RCW~38 has $\sim$50-times more stars than expected for a cluster its size\footnote{A recent analysis of RCW~38 by \citet{2017MNRAS.471.3699M} has reported a maximum surface density $\sim$6 times lower than reported in MYStIX using near-infrared observations. This difference may arise due to correction for completeness or differences in data-smoothing method to calculate density. The \citet{2017MNRAS.471.3699M} central density moves this point closer to the regression line found for the other MYStIX clusters; however, it would still be nearly an order of magnitude more dense than predicted by the regression line.} given the $N_4\sim r_4$ regression line, and its unusually high central density has already been noted by \citet{2015ApJ...802...60K}.

These observations would suggest that NGC~6231 arose from the same cluster assembly processes that formed the majority of MYStIX subclusters because a different cluster formation process would not necessarily produce a cluster following the same age--radius--mass--density relations. Below we briefly describe the subcluster relations obtained by \citet{2015ApJ...812..131K} and how NGC~6231 relates to each case.
\begin{description}
\item[Radius--age relation] The regression line\footnote{Several methods exist to obtain linear regression fits to bivariate data. Here we present the slope of the reduced major-axis regression line to the data on a log-log plot. In contrast, \citet{2015ApJ...802...60K} provide the orthogonal regressions, but these are mislabeled as reduced major-axis regressions in their Tables 3 and 4.} shown for MYStIX subclusters is $r_4 \propto age^{1.8}$. The statistically significant correlation between age and radius was interpreted as cluster expansion, which may be an effect of mass loss (e.g., loss of cloud material), binary stars, subcluster mergers, or a cluster that is initially supervirial or unbound. For an age of 3.2~Myr, NGC~6231 is slightly larger than given by the regression, but well within the scatter observed for MYStIX subclusters. A 6.4~Myr age would place NGC~6231 below the regression line.
\item[Number of stars--radius relation] The regression line shown is $N_4 \propto r_4^{1.4}$. Several effects could produce this relation, including build up of cluster mass while expansion is occurring as suggested by \citet{2015ApJ...812..131K} or a birth relation inherited from the mass--size relation of molecular clumps as suggested by \citet{2016A&A...586A..68P}. The coincidence of NGC~6231 with the regression line is quite close. However, given that NGC~6231 has almost certainly expanded from its original size, the second explanation is unlikely in this case.
\item[Density--radius relation] The regression lines shown are $\rho_0 \propto r_4^{-2.4}$ and  $\Sigma_0 \propto r_4^{-1.6}$. The slopes of these lines are slightly less steep than the relation that would be expected for a cluster expanding while neither gaining nor losing stars (i.e. $\rho_0 \propto r_4^{-3}$).  \citet{2015ApJ...812..131K} proposed that a cluster gaining stars through hierarchical mergers of subclusters could produce such a relation. Alternatively, the distribution could be produced by combined effects of initial conditions and evolution of the cluster. Although NGC~6231 is less dense than most MYStIX subclusters, it has a higher density than expected given its radius. 
\end{description}

It is intriguing that a simple, isolated, massive cluster like NGC~6231 would follow the same relations as less massive subclusters in complex regions with ongoing star formation. If this is not a coincidence, then it may suggest that the same processes that govern the properties of subclusters in star-forming complexes also govern the properties of fully-formed young clusters. Many of the MYStIX star-forming regions are likely sites of hierarchical cluster assembly. \citet{2009MNRAS.397..954F} show that subclusters still embedded in clouds (like MYStIX regions DR~21 or NGC~6334) rapidly merge. \citet{2014ApJ...787..107K} note that the variation in morphology of young star clusters (ranging from linear chains of subclusters, clumpy distributions of stars, centrally concentrated clusters) may be an evolutionary progression. Thus, we suggest that the common factor between NGC~6231 and MYStIX subclusters is hierarchical assembly. 

\subsection{Current State}

The CXOVVV catalog can reveal the effects of the cluster's dynamical state on its spatial structure, but kinematic data are necessary to obtain fundamental properties of the current state, such as the cluster's total energy and how the energy is partitioned. In the near future, measurements of proper motion are expected to become available from the Gaia survey. For the stars in the CXOVVV catalog, it is estimated that the precision will be 0.1--2.0~km~s$^{-1}$, although the performance may be degraded near the Galactic plane \citep{deBruijne2005gaia}. These measurements could be used to test some of the inferences about the cluster's current state from this paper.

\subsubsection{Radial Structure}\label{radial.sec}

The projected density of stars in NGC~6231, based on star counts in the CXOVVV catalog, is one of the most accurately known of nearby young star clusters. This has allowed us to test a variety of empirical cluster models. We find that the isothermal ellipsoid model is a remarkably good fit, while other commonly used models are not. We note that the isothermal ellipsoid is a special case of the EFF profile \citep{1987ApJ...323...54E}, which has been found to describe the distribution of light in very massive young clusters in the Milky Way and nearby galaxies. Recently, \citet{2017arXiv170809065G} have argued that EFF surface-density profile arises from hierarchical cluster assembly. 

Another theoretical implication of this result is that the formation of star clusters is not an entirely scale-invariant process, given that the resulting cluster does have a characteristic length scale, $r_c$. This is interesting because many of the astrophysical processes of star formation are scale invariant, leading to fractal-like distributions of star-formation activity in the Galaxy. Nevertheless, within individual star clusters, some process must produce the observed length scales.

\subsubsection{Cluster Expansion}\label{expansion.sec}

NGC~6231's current core radius of 1.2$\pm$0.1~pc is a factor of $\sim$15 larger than the typical core radius of highly-absorbed subclusters of stars in MYStIX, and a factor of $\sim$7 larger than the typical MYStIX young stellar cluster \citep{2014ApJ...787..107K}. Other studies also suggest that typical embedded clusters have half-mass radii of a few tenths of parsecs \citep[e.g.,][]{2012A&A...543A...8M}. Thus, NGC~6231's size is likely dominated by expansion, not its initial formation size.

The cause of cluster expansion can have an effect on a cluster's radial structure, including the surface-density profile and the presence or absence of a radial--age gradient. Expansion of young clusters can be driven by mass loss from the dispersal of a molecular cloud, while older clusters can expand due to mass loss from stellar evolution. However, a gravitationally bound cluster will expand even in the absence of these effects, and \citet{2012MNRAS.426L..11G} has suggested that this expansion can explain the surface-density distribution of pre-main-sequence stars in our Galactic neighborhood. This expansion is driven by transfer of energy from binaries, mass segregation, and mass loss due to dynamical relaxation, and the resulting expansion is homologous (preserving radial structure) and scale-invariant in time \citep{1996MNRAS.279.1037G}. 

In contrast, an unbound cluster may have a different structure. If the total energy of a cluster is highly positive, the cluster will evolve toward a structure determined by the velocity dispersion of its stars. If stars have a Maxwell-Boltzmann distribution, this would be a multivariate normal surface density distribution. Alternatively, in a cluster where a fraction of stars are escaping, models by \citet{2006MNRAS.369L...9B}  show that the escaping stars may lead to an excess number of stars at large radii compared to an EFF profile. Given that NGC~6231 is well described by the isothermal ellipsoid model (Figure~\ref{cuts.fig}) and does not appear to have excess stars within the {\it Chandra} field of view, it is unlikely that the cluster disintegrating in either of these ways. 

The test for a radial gradient in stellar ages can also be used to evaluate whether stars have been escaping from the region from the cluster during the star-formation process. If stars in NGC~6231 were free to drift from their initial location as soon as they were formed, one would expect the first stars that formed to have drifted the farthest, creating a radial--age gradient. The age spread in NGC~6231 has been estimated to be $\Delta age\sim2$--7~Myr \citep{2007MNRAS.377..945S,2013AJ....145...37S,2016arXiv160708860D}, and most stellar ages calculated in \citetalias{KMG17} range from 2.5~Myr to 9~Myr. If stars have drifted outward over this period, the oldest stars would have on average moved farthest from the center of the cluster, producing a clear radial-age gradient. However, no radial age gradient is seen, and the stars of various ages are well mixed. Thus, the cluster must have been gravitationally bound at least until star formation had ceased. In contrast, we note that the MYStIX cluster NGC~2244 does have hints of a radial--age gradient in the analysis of \citet[][their Figure~7c]{2014ApJ...787..108G}.

\subsubsection{Dynamical Evolution} 

Dynamical timescales are important for cluster evolution. However, the lack of kinematic data for NGC~6231 means that the timescales can only be approximated. Velocity dispersions are likely to be similar to those in other young clusters, which range from $\sim$1--3~km~s$^{-1}$ \citep[e.g.,][]{2008ApJ...676.1109F,2012A&A...539A...5C,2016A&A...588A.123R}. We also use $r_4=4.6$~pc as a characteristic radius for the cluster, yielding a cluster crossing time of $t_\mathrm{cross}=1.5$--4.6~Myr. The number of stars within this three-dimensional volume\footnote{For these calculations we will ignore the effects of cluster elongation, treating the cluster as a radially symmetric distribution of stars.} is $N\approx4300$~stars (\S\ref{tot_mass.sec}). The timescale for cluster virialization through two-body interactions is approximated by \citet{2008gady.book.....B} as
 \begin{equation}
t_\mathrm{vir} = t_\mathrm{cross}\times\frac{N}{8\ln N}.
\end{equation}
Thus, the two-body virilization timescale for NGC~6231 would be $\sim$60 cluster crossing times, or $\sim$100--300~Myr, much longer than the cluster has existed. Nevertheless, in the cluster's evolution so far, it is likely that more rapid dynamical processes (e.g., violent relaxation) have been dominant as we discuss below.  

Some structural properties of NGC~6231 suggest that the cluster has undergone some dynamical evolution. This includes a surface density distribution that has a radial profile similar to what is expected from a kinematically isothermal cluster, the segregation of high-mass stars, and the thorough mixing of stars of different ages. On the other hand, some of the observed features of NGC~6231 would likely have been erased in a cluster that had already reached a quasi-equilibrium state: these include the existence of a small subcluster and the possible asymmetry in the main cluster mentioned in Section~\ref{single.sec}. Subclustered structure would be erased during dynamical relaxation, and \citet{1998ApJ...492..540H} and \citet{2005ApJS..160..379F} attribute an asymmetry in the spatial configuration of the Orion Nebula Cluster to ongoing violent relaxation.  However, it is also possible that the subcluster is physically separated from NGC~6231 and is just projected onto the main cluster by chance. The age of the NGC~6231 of 2--7~Myr is much less than the 100--300-Myr dynamical-relaxation timescale.

The spatial distribution of stars in NGC~6231 appears to be smoother on shorter length-scales (within the {\it Chandra} field of view) and more clumpy on larger length-scales (outside the {\it Chandra} field of view). This may be related to different cluster crossing times and dynamical timescales for different length scales. For example, the cluster crossing time of 1.5--4.6~Myr within a radius of $r_4$ is less than or approximately equal to the median age of stars in the cluster. While, the cluster-crossing timescale outside this region, where the distribution is clumpy can be significantly larger than the median age of the stellar population. However, there is not sufficient information to definitively distinguish subclusters of stars in the Sco~OB1 complex from unrelated young star clusters or associations along the same field of view

\subsubsection{Mass Segregation}
Mass segregation has been observed in a number of young star clusters. For example, mass segregation is seen for high-mass stars in some star-forming regions \citep[e.g., the Orion Nebula Cluster;][]{1998ApJ...492..540H} and low-mass stars in others \citep[e.g., W40;][]{2010ApJ...725.2485K}. A variety of astrophysical phenomena can produce mass segregation, making it difficult to test any particular model. For example, mass segregation will appear in dynamically relaxed clusters due to two-body interactions in which high-mass stars lose energy and sink toward the centers of clusters. The theoretical models for this process are complicated. Several families of models have been discussed by \citet{1961AnAp...24..369H}, \citet{1979AJ.....84..752G}, and \citet{2015MNRAS.454..576G}. However, mass segregation can also be induced rapidly (in a single cluster crossing time) in young star clusters through violent relaxation or mergers of subclusters \citep{2007ApJ...655L..45M,2009ApJ...700L..99A,2010MNRAS.404..721M}. It may also be the case that OB stars can only form in certain regions, leading to primordial mass segregation. \citet{2014MNRAS.438..620P} found that dynamical mass segregation will increase with dynamical age as clusters become more radially symmetric.

Although OB stars in NGC~6231 are mass segregated, the mass segregation in this cluster is not as strong as in other star-forming regions. The differences in radial distribution between massive stars and low-mass stars (Figure~\ref{mass_seg2.fig}) are not as prominent as in the Orion Nebula Cluster \citep[Figure~6 from][]{1998ApJ...492..540H}, and the segregation does not extend down to low-mass stars as it does in W40. On the other hand, some other young star clusters do not show mass segregation \citep[e.g., NGC~2244;][]{2008ApJ...675..464W}. Both NGC~6231 and NGC~2244 are older than Orion and W40, suggesting that mass segregation in newly formed clusters may not be just a simple effect of dynamical age. 

Mass segregation can arise from the dissintegration of an unbound cluster if stellar velocities depend on mass because, after a stellar association had expanded sufficiently, the distance of a star from the center of the cluster would be proportional to its velocity. If such a cluster had energy equipartition ($\sigma_v(m)\propto m^{-1/2}$), this would result in a power-law relation with index $-1/2$ between stellar mass and characteristic radii. Figure~\ref{equipartition.fig} shows that the regression line for core radius versus average mass is slightly less steep than a  $-1/2$ relation.

\subsubsection{Comparison to Very Massive Young Clusters}\label{pz.sec}

\begin{figure}
\centering
\includegraphics[width=0.45\textwidth]{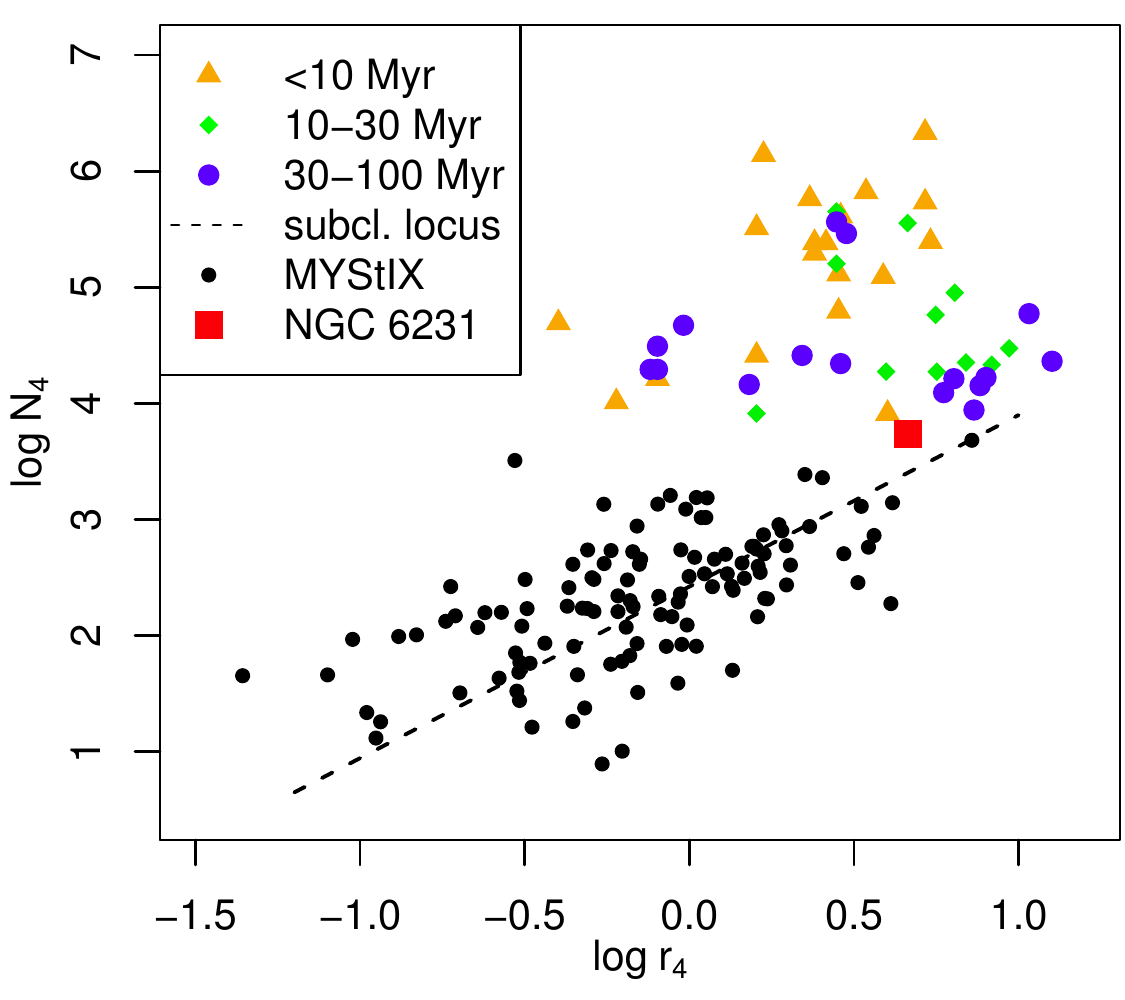} 
\caption{Comparison of NGC~6231 to massive young clusters from \citet{2010ARA&A..48..431P} (both Galactic and extragalactic) and MYStIX subclusters.  NGC~6231 is marked by the red square, while the massive young clusters are marked by symbols indicating their age (orange triangles, green diamonds, and blue circles), and the MYStIX subclusters are shown as black circles. The dashed black line shows the regression line found for MYStIX subclusters from Figure~\ref{mystix_comparison.fig}. Only clusters from the \citet{2010ARA&A..48..431P} sample with estimates of core radius are included. \label{mysc.fig}}
\end{figure}

Insight into cluster formation mechanisms can also be gained by comparing NGC~6231 to very massive young star clusters. The review by \citet{2010ARA&A..48..431P} gives a sample that includes several of the most extreme young star clusters ($M\sim10^4$--$10^7$~$M_\odot$) in the Milky Way and nearby galaxies, hereafter the PZ sample. In contrast, NGC~6231 and the MYStIX star-forming regions are all in our relatively quiet neighborhood of the Galactic Disk; neither the starburst at the Galactic Center nor where the Galactic Bar ends. Figure~\ref{mysc.fig} shows NGC~6231, the MYStIX subclusters, and the PZ clusters on the same plot of radius vs.\ number of stars.\footnote{There is one cluster in common in the two samples: Tr~14. The PZ properties of this cluster come from \citet[][]{2007A&A...476..199A}. The measured central densities of stars in the clusters are similar, $7.3\times10^3$~stars/pc$^3$ (PZ) vs.\ 10$^4$~stars/pc$^3$ (MYStIX), and the measured core radii are similar, 0.14~pc (PZ) vs. 0.2~pc (MYStIX). Nevertheless, the total number of stars reported by \citet{2007A&A...476..199A} is $\sim$5 times greater than reported by MYStIX. The explanation for this is that \citet{2007A&A...476..199A} include a much larger fraction of the young stellar population of the Carina Nebula Cluster as part of Tr~14 than is included in MYStIX. Thus, we only include the Tr~14 measurement from MYStIX; however, the reader is cautioned that similar differences in definition of cluster boundary may affect other starburst clusters from the PZ sample.} NGC~6231 lies between these samples in mass, more massive than the MYStIX subclusters and less massive than the PZ clusters. The properties compiled for the  PZ clusters include a core radius $r_c$ where surface density decreases by a factor of $\sim$2 from the central density, an effective radius $r_\mathrm{eff}$ that contains half the light from the cluster, and a photometric cluster mass. The core radius is directly comparable to our values of $r_c$ for NGC~6231 and the MYStIX subclusters. We do not have a direct measurement of total mass for NGC~6231 or the MYStIX subclusters. However, for the PZ sample, we can approximate $N_4$ by assuming an average stellar mass of $\bar{m}=0.61$~$M_\odot$ and assuming that approximately half the cluster mass is contained within a radius $r_4=4r_c$. (Given that Figure~\ref{mysc.fig} is a log-log plot, errors resulting from these approximations likely have little effect on the overall distributions of points.)

NGC~6231 has fewer stars than every starburst clusters, as tabulated by \citet{2010ARA&A..48..431P}. However, NGC~6231 also has a larger core radius than most of them, including all of the PZ clusters with estimated ages less younger than NGC~6231, and 16 out of 19 PZ clusters with ages less than 10~Myr. This may suggest that NGC~6231 has expanded more than the PZ clusters. NGC~6231 has a lower central density than all but a few of the oldest clusters in the PZ sample.  

From Figure~\ref{mysc.fig}, it can be seen that the MYStIX mass--radius relation, which appears to hold for NGC~6231, does not hold for massive young star clusters in general. For the PZ clusters, the range of core-radius values is similar to that of the larger MYStIX clusters and subclusters, but their $N_4$ values are greater by 2 to 4 orders of magnitude, which places them all above the MYStIX mass--radius relation. A subset of $\sim$12 PZ clusters do lie on the $r_4$--$N_4$ plot to the upper right of NGC~6231, and these are only slightly off the MYStIX mass--radius relation, but these are a minority of the sample. 

The different locations of very massive clusters and less massive clusters/associations represented by MYStIX and NGC~6231 on the $r_4$--$N_4$ plot suggests different mechanisms of cluster assembly and/or different types of cluster evolution. For example, \citet{2014ApJ...787..158B,2015MNRAS.447..728B} argue for monolithic (or prompt) formation of the massive young cluster R136 (left-most orange triangle). \citet{2009A&A...498L..37P,2011A&A...536A..90P} suggest distinct cluster evolution for ``starburst'' versus ``leaky'' young star clusters. NGC~6231 and the MYStIX subclusters have masses of the more-common leaky clusters, while the PZ clusters have masses of the less-common starburst clusters. 

\subsection{Fate of NGC~6231}

The fate of NGC~6231 is tied to the issue of gravitational boundedness. If the system is unbound, the members may form a coherent moving group for a number of Galactic orbits \citep[e.g.,][]{2005A&A...435..875F}, but the surface density will rapidly diminish to the point where the it is nearly indistinguishable from the field \citep{2015A&A...576A..28P}. 
We have provided evidence that stars were born gravitationally bound to the system, but that the cluster has likely expanded by a factor of 7--15 since its birth. Here, we investigate whether this expansion indicates that the cluster is already unbound or liable to tidal disruption.   

\subsubsection{Gravitational Boundedness or Unboundness \label{bound.sec}}

\begin{figure}
\centering
\includegraphics[width=0.45\textwidth]{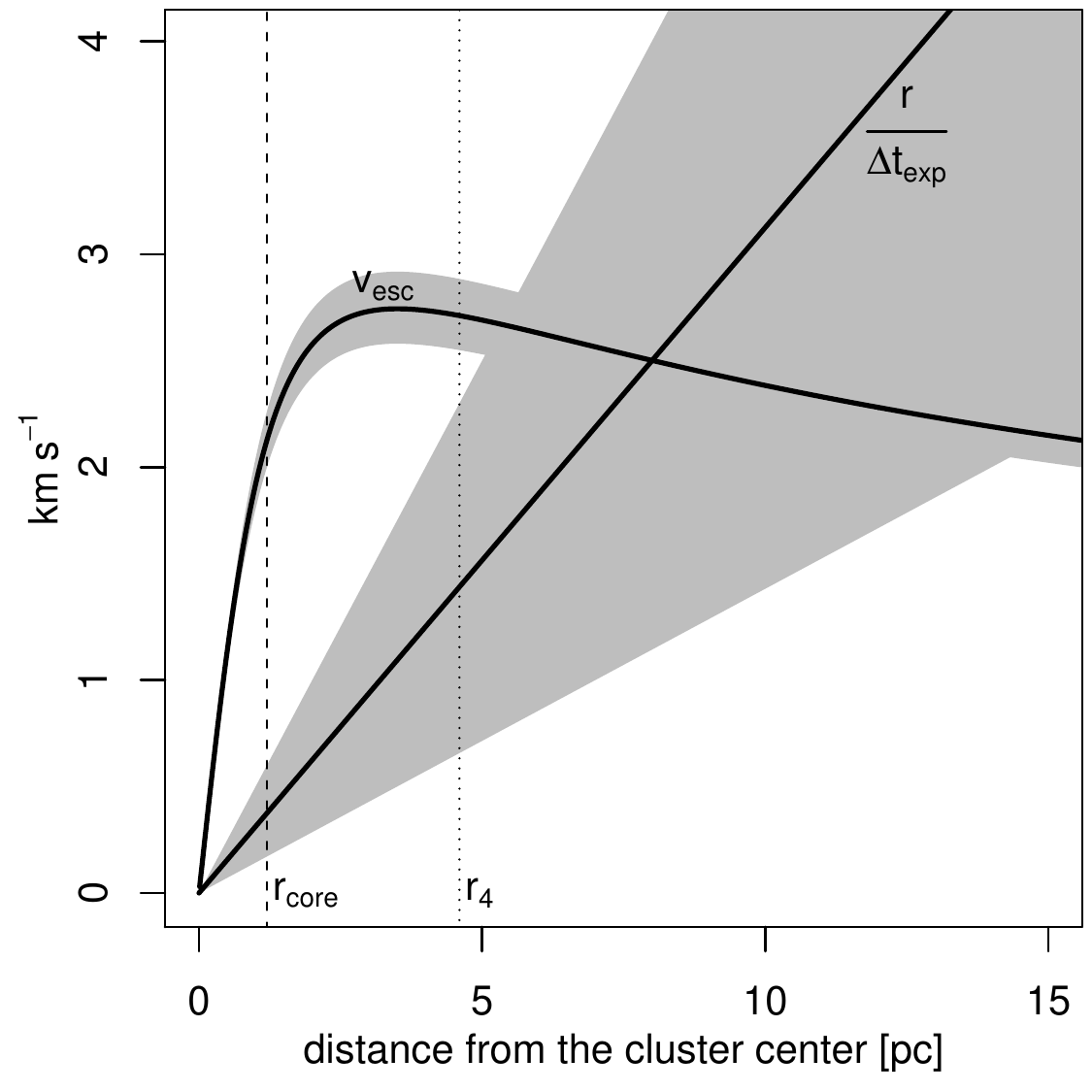} 
\caption{The curved black line shows escape velocity $v_\mathrm{esc}$ as a function of distance from the cluster center $r$, ignoring stars outside radius $r$. The straight black line shows the velocity that a star would need to travel from the center of the cluster to distance $r$ in the time $\Delta t_{exp}$ since the cluster started to expand. When $r / \Delta t_{exp} > v_\mathrm{esc}$, stars are not gravitationally bound, assuming they originated near the cluster center. Since this is only the case at large radii, greater than the cluster core radius $r_c$ or the cluster characteristic radius $r_4$, it is likely that most stars in the cluster are gravitationally bound. We assume a cluster central density of 200~stars~pc$^{-3}$ and a cluster age of 3.2~Myr, but the gray regions show uncertainties in these lines due to the uncertainty in cluster mass and an age range of 2--7~Myr. 
 \label{escape.fig}}
\end{figure}

We can test whether the velocity of stars required to expand NGC~6231 from an initially compact configuration to its current size is greater than the escape velocity. To calculate this velocity, we assume that velocities are mostly radial, which is a reasonable assumption if stars are escaping. For this discussion, we use the mean velocity of a star from its point of origin to its current location, but a star with an outward trajectory would typically be slowing. We also assume that if stars at central distance $r$ are unbound, then stars at larger radii will also be unbound. Integrating the mass within a radius $r$ for the three-dimensional Hubble model gives the equation for the escape velocity,
\begin{equation}
v_\mathrm{esc} = \sqrt{\frac{8\pi G r_c^3 \rho_0\bar{m}}{r}\left(\sinh^{-1}(r/r_c) - \frac{r/r_c}{(r^2/r_c^2+1)^{1/2}} \right)},
\end{equation}
where $G$ is the gravitational constant. Cluster expansion most likely started 2--7~Myr ago, so the mean velocity of a star at distance $r$ from the cluster center would be $v_\mathbf{mean}=r/(2~\mathrm{Myr})$ to $r/(7~\mathrm{Myr})$. Figure~\ref{escape.fig} shows a comparison of the escape velocity curve and to the mean velocity a star would need to travel a distance $r$ since the cluster expansion began. The gray regions show the effect of uncertainty on cluster mass and uncertainty on when cluster expansion began, which we assume to be approximately equal to the age of the cluster. If $\rho_0=200$~stars~pc$^{-3}$ and expansion started 3.2~Myr ago, the average outward velocity would be lower than the escape velocity out to $\sim$8~pc, much larger than the $r_4$ radius. The peak escape velocity would be 2.75~km~s$^{-1}$. 

The expansion of a cluster core to $\sim$1~pc during the first several million years is also not inconsistent with simulations of a bound cluster. For example, \citet{2001MNRAS.321..699K} and \citet{2006MNRAS.373..752G} find expansion of this magnitude in simulations of bound clusters, with the former describing the simulation as a ``Pleiades'' analog.

\subsubsection{Tidal Effects}

A young star cluster must be smaller than its Jacobi radius if it is to survive as a open cluster in the Milky Way without disruption due to tidal stripping. The Jacobi radius is 
\begin{equation}
r_J = \sqrt[3]{\frac{GM}{4A(A-B)}},
\end{equation}
where $M$ is cluster mass, $G$ is the gravitational constant, and $A$ and $B$ are Oort's constants at the location of the cluster \citep[][his Equation~24]{1962AJ.....67..471K}. The values of the Oort constants at the location of NGC~6231 are $A\approx11.8$~km~s$^{-1}$ and $A-B\approx 22.2$~km~s$^{-1}$ using the equations from \citet{2006A&A...445..545P,2007A&A...468..151P}. For a total cluster mass of 2200--6800~$M_\odot$, the Jacobi radius would be $r_J=20$--30~pc, or $0.\!\!^\circ7$--$1.\!\!^\circ1$ on the sky. Given that $\ge$50\% of stars reside with a radius $r_4=4.6$~pc from the cluster center, NGC~6231 is currently much smaller than its Jacobi radius. 

Young clusters may also be tidally disrupted by molecular clouds and other young star clusters in their natal environment, in a phenomenon known as the ``cruel cradle effect'' \citep{2012MNRAS.426.3008K}. The remaining massive elements of Sco~OB1, besides NGC~6231, include the clouds IC~4628 and the Large Elephant Trunk and the clusters Tr~24, VDBH~211, 207, G342.1+00.9, and various other groupings of stars discussed in \S\ref{largescale.sec}. IC~4628 is likely to be the most massive of these, with a total mass of more than $10^3$~$M_\odot$ \citep{1986A&AS...65..465P}, but more precise mass estimates are not available from the literature. Nevertheless, at an angular distance of $1.\!\!^\circ9$ from NGC~6231, IC~4628 would need to be at least $\sim$20 times more massive than NGC~6231 to tidally disrupt the cluster. 





\section{Conclusions}\label{conclusions.sec}

The new CXOVVV catalog of 2148 probable cluster members in NGC~6231 from \citetalias{KMG17} can be used as a testbed for star-cluster formation theory. This sample is only complete down to 0.5~$M_\odot$; the total intrinsic stellar population of cluster members within the {\it Chandra} field of view is estimated to be $\sim$5700 stars (\S 2-3), and the full population is likely to be significantly larger. 

The isothermal ellipsoid (Equation~\ref{hubble.eq}) provides a remarkably good empirical fit to the surface density distribution of stars in the cluster. This model is notably better than Plummer sphere, multivariate normal, or power-law models. The cluster has core radius $r_c=1.2\pm0.1$~pc and ellipticity $\epsilon=0.33\pm0.05$. 4\% of the stars are in a small subcluster embedded in, or projected upon, the main cluster (\S 5). Several additional small young clusters are present within 30~pc within the large Sco~OB1 complex (\S 8). Mass segregation is present with a statistical significance of $p\sim0.001$ for $M>8$~$M_\odot$ and lower significance for lower-mass stars (\S 6). The empirical dependence of spatial dispersion stellar mass is given by $r_c\propto M^{-0.29}$. The median age of stars in the cluster is around 3-4 Myr with a significant age spread, but no radial age gradient is present (\S 7). The basic structural properties of the cluster measured in this paper are listed in more detail in \S\ref{summary.sec} of the Discussion.

NGC~6231 has physical properties similar to other clusters and association in the nearby Galaxy studied in the MYStIX project. NGC~6231 most closely resembles NGC~2244 in the Rosette Nebula. Both clusters are similar in size, number of stars, and age, and both are located in bubbles from which gas has been removed. In addition, NGC~6231 lies on the empirical relationships between core radius, age, mass, and density found for clusters and subclusters in the MYStIX star-forming regions (\S\ref{assembly.sec}). This could be considered to be an unexpected result, given that the spatial distributions of stars in MYStIX star-forming regions with clumpy, subclustered structures (like NGC~6334 or the Carina Nebula) are very different from the smooth distribution of stars in NGC~6231. However, this situation may arise if both subclusters in star-forming regions and more evolved clusters like NGC~6231 are assembled through similar processes. In contrast, extremely massive young star clusters in the Milky Way and nearby galaxies from the PZ sample do not follow this relation, possibly suggesting a different mode of cluster formation for very massive clusters (\S\ref{pz.sec}).

Several lines of evidence indicate that NGC~6231 is gravitationally bound. The lack of a radial gradient in stellar age shows that most stars have not been freely drifting away from the cluster during the several million years (approximately 2 to 7~Myr ago) when stars were forming (\S\ref{radial.sec}). 
The cluster's expansion velocity is significantly less than the escape velocity for stars, which would not be true for an unbound cluster. The cluster is also significantly smaller than its tidal disruption radius (\S\ref{bound.sec}). 

Future measurements of cluster kinematics will be able to better constrain theoretical modeling. Proper motions from Gaia are expected to have precisions of 0.1--2~km~s$^{-1}$. When combined with object selection using the CXOVVV catalog, the Gaia dataset will provide four-dimensional $(x,y,v_x,v_y)$ kinematic information about the cluster, which will allow the conclusions from this paper, based on the two-dimensional spatial distributions of stars, to be tested.

\acknowledgments MAK, EDF, MG, NM, JB, and RK acknowledge support from the Ministry of Economy, Development, and Tourism's Millennium Science Initiative through grant IC120009, awarded to The Millennium Institute of Astrophysics. MAK was also supported by a fellowship (FONDECYT Proyecto No.\ 3150319) from the Chilean Comisi\'on Nacional de Investigaci\'on Cient\'ifica y Tecnol\'ogica, and RK received support from FONDECYT Proyecto No.\ 1130140. The scientific results reported in this article are based on data obtained from the Chandra Data Archive, the Vista Variables in V\'ia Lact\'ea project (ESO program ID 179.B-2002), and the Two Micron All Sky Survey catalog. This work make use of analysis methods developed at Penn State for the MYStIX project and {\it Chandra} data reduction procedures (including ACIS Extract) developed by Patrick Broos and Leisa Townsley. We thank the referee for many useful comments and suggestions. We also thank Anushree Sengupta for useful feedback on the article and Amelia Bayo and Estelle Moraux for helpful discussions about star-forming regions.

\facility{CXO(ACIS-I), ESO:VISTA(VVV)}

\software{
          ACIS Extract \citep{2010ApJ...714.1582B},
          astro \citep{CRANastro},
          astrolib \citep{1993ASPC...52..246L},  
          astrolibR \citep{CRANastrolibR},
          celestial \citep{2016ascl.soft02011R},
          SAOImage DS9 \citep{2003ASPC..295..489J}, 
          SIMBAD \citep{2000A&AS..143....9W},
          spatstat \citep{baddeley2005spatstat}, 
          TOPCAT \citep{2005ASPC..347...29T},
          XPHOT \citep{2010ApJ...708.1760G}
          }

\bibliography{mybib}



\end{document}